%% file: PolarSkyrmionCombined.tex
\documentclass[pdflatex,sn-mathphys]{sn-jnl}

\jyear{2021}%

\theoremstyle{thmstyleone}%
\theoremstyle{thmstyletwo}%

\theoremstyle{thmstylethree}%

\usepackage{amsmath}
    {\end{bmatrix}\end{medsize}}%

\usepackage[left]{lineno}

\usepackage[normalem]{ulem}

\usepackage{etoolbox}
\makeatletter
\patchcmd{\ps@headings}
{\hbox to \hsize{\hfill Springer Nature 2021 \LaTeX\ template\hfill}}
{\hbox to \hsize{}}
{}
{}
\patchcmd{\ps@headings}
{\hbox to \hsize{\hfill Springer Nature 2021 \LaTeX\ template\hfill}}
{\hbox to \hsize{}}
{}
{}
\patchcmd{\ps@titlepage}
{\hbox to \hsize{\hfill Springer Nature 2021 \LaTeX\ template\hfill}}
{\hbox to \hsize{}}
{}
{}
\makeatother

\usepackage{caption}
\begin{document}

\title[~]{\vspace{-2cm} Emergent higher-order optical nonlinearity in self-assembled polar supertextures}

\author[1]{\fnm{Daniel A.} \sur{Bustamante Lopez}}\email{dabl@bu.edu}
\author[1]{\fnm{Deepankar} \sur{Sri Gyan}}\email{dsrigyan@bu.edu}

\author[2]{\fnm{Xiaojiang} \sur{Li}}\email{xiaojiangli0117@gmail.com}
\author[2]{\fnm{Vladimir A.} \sur{Stoica}}\email{vxs30@psu.edu}

\author[3]{\fnm{Sujit} \sur{Das}}\email{sujitdas@iisc.ac.in}

\author[4]{\fnm{Peter} \sur{Meisenheimer}}\email{peter@keplercompute.com}
\author[4]{\fnm{Akash} \sur{Surampalli}}\email{as580@rice.edu}

\author[3] {\fnm{Mohit} \sur{Tanwani}}\email{mohittanwani@iisc.ac.in}

\author[5,6]{\fnm{Xianghan} \sur{Xu}}\email{xu001395@umn.edu}

\author[2]{\fnm{Aiden} \sur{Ross}}\email{amr8057@psu.edu}
\author[2]{\fnm{Long-Qing} \sur{Chen}}\email{lqc3@psu.edu}

\author[5]{\fnm{Sang-Wook} \sur{Cheong}}\email{sangc@physics.rutgers.edu}
\author[7,8,9,10,11]{\fnm{Lane W.} \sur{Martin}}\email{lwmartin@rice.edu}
\author[7,8,10,11,12,13]{\fnm{Ramamoorthy} \sur{Ramesh}}\email{ramamoorthy.ramesh@rice.edu}

\author[2]{\fnm{Venkatraman} \sur{Gopalan}}\email{vxg8@psu.edu}
\author*[1,14,15]{\fnm{Wanzheng} \sur{Hu}}\email{wanzheng@bu.edu}

\affil[1]{\orgdiv{Department of Physics}, \orgname{Boston University}, \orgaddress{\street{590 Commonwealth Avenue}, \city{Boston}, \postcode{02215}, \state{Massachusetts}, \country{USA}}}

\affil[2]{\orgdiv{Department of Materials Science and Engineering}, \orgname{Pennsylvania State University}, \orgaddress{\street{University Park}, \postcode{16802}, \state{Pennsylvania}, \country{USA}}}

\affil[3]{\orgdiv{Materials Research Centre}, \orgname{Indian Institute of Science}, \orgaddress{ \postcode{560012}, \city{Karnataka}, \country{India}}}

\affil[4]{\orgdiv{Department of Materials Science and Engineering}, \orgname{University of California, Berkeley}, \orgaddress{ \postcode{94720}, \city{California}, \country{USA}}}

\affil[5]{\orgdiv{Keck Center for Quantum Magnetism and Department of Physics and Astronomy}, \orgname{Rutgers University}, \orgaddress{\street{136 Frelinghuysen Road}, \city{Piscataway}, \postcode{08854}, \state{New Jersey}, \country{USA}}}

\affil[6]{\orgdiv{Present address: School of Physics Astronomy}, \orgname{University of Minnesota}, \orgaddress{\street{115 Union St. SE}, \city{Minneapolis}, \postcode{55455}, \state{Minnesota}, \country{USA}}}

\affil[7]{\orgdiv{Department of Materials Science and Nanoengineering}, \orgname{Rice University}, \orgaddress{ \postcode{77005}, \city{Houston}, \state{Texas}, \country{USA}}}
\affil[8]{\orgdiv{Department of Physics and Astronomy}, \orgname{Rice University}, \orgaddress{ \postcode{77005}, \city{Houston}, \state{Texas}, \country{USA}}}
\affil[9]{\orgdiv{Department of Chemistry}, \orgname{Rice University}, \orgaddress{ \postcode{77005}, \city{Houston}, \state{Texas}, \country{USA}}}
\affil[10]{\orgdiv{Rice Advanced Materials Institute}, \orgname{Rice University}, \orgaddress{ \postcode{77005}, \city{Houston}, \state{Texas}, \country{USA}}}

\affil[11]{\orgdiv{Materials Sciences Division}, \orgname{Lawrence Berkeley National Laboratory}, \orgaddress{ \postcode{94720}, \city{Berkeley}, \state{CA}, \country{USA}}}

\affil[12]{\orgdiv{Department of Materials Science and Engineering}, \orgname{University of California, Berkeley}, \orgaddress{ \postcode{94720}, \city{Berkeley}, \state{CA}, \country{USA}}}
\affil[13]{\orgdiv{Department of Physics}, \orgname{University of California, Berkeley}, \orgaddress{ \postcode{94720}, \city{Berkeley}, \state{CA}, \country{USA}}}
\affil[14]{\orgdiv{Division of Materials Science and Engineering}, \orgname{Boston University}, \orgaddress{\street{590 Commonwealth Avenue}, \city{Boston}, \postcode{02215}, \state{Massachusetts}, \country{USA}}}

\affil[15]{\orgdiv{Photonics Center}, \orgname{Boston University}, \orgaddress{\street{8 Saint Mary's St.}, \city{Boston}, \postcode{02215}, \state{Massachusetts}, \country{USA}}}

\abstract{Giant optical nonlinearities enable advanced materials control and bridge exotic light-matter phenomena to next-generation photonics. However, materials exhibiting strong higher-order optical nonlinearity beyond the second order are typically rare. Here, we report the discovery of a giant and ultrafast second-harmonic generation (SHG) in self-assembled polar skyrmions under mid-infrared excitations. The transient SHG intensity exhibits a distinctive power dependence originating from higher-order nonlinearity. More generally, we show that large higher-order nonlinearity emerges across diverse domain structures in polar superlattices. Phase-field simulations demonstrate a strong and periodic local permittivity variation within each polar texture at equilibrium, and further confirm the fast polarization oscillation in polar skyrmions with mid-infrared excitations. This intense local permittivity modulation can satisfy epsilon-near-zero conditions, unlocking dramatically enhanced nonlinear optical responses without photonic engineering. Consequently, the self-assembled polar topologies with strong inherent optical nonlinearity could establish a platform for quantum information processing and high-speed optical communications.}


\maketitle

\section*{Introduction}\label{sec1}

Optical nonlinearity\cite{YRShenBook,BoydBook} has revolutionized material control since the advent of lasers. Materials with strong optical nonlinearities, such as non-centrosymmetric crystals, plasmonic nanostructures, and engineered metamaterials are critical for nonlinear optical applications\cite{Fiebig2005,Denev2011,Kauranen2012,Pettine2024}. Second-order nonlinear effects drive frequency doubling\cite{Franken1961,Sionnest1987,Manzoni2016}. Third-order effects enable four-wave mixing including third harmonic generation\cite{New1967,Hendry2010,Yue2024}. Higher-order nonlinear optical phenomena (third order and above) in conventional materials are typically weak\cite{Shkurinov1993,Hafez2018}, limiting their utility in integrated photonics. Materials exhibiting strong higher-order optical nonlinearities with short response times are highly desirable for optoelectronic devices that leverage multiphoton interactions and non-degenerate wave mixing processes\cite{Yue2024}. Such devices, including photonic switches and frequency converters, are critical for applications in quantum information processing, optical remote sensing, and high-speed optical communications\cite{Dutt2024}. However, achieving these properties often requires complex and intricate material engineering. In artificially structured metamaterials and photonic crystals, the strong spatial variation of permittivity from periodically arranged materials enables the epsilon-near-zero condition, thereby greatly enhancing nonlinear optical response\cite{Liberal2017}. Alternatively, nanoscale self-assembled polar textures with large optical nonlinearity offer a new route for nanophotonics design\cite{Sun2025}.

Ferroelectric superlattices have emerged as engineered quantum materials wherein layers of ferroelectric and dielectric materials are stacked periodically at the nanoscale. In superlattices, mechanical constraints, particularly epitaxial strain from lattice mismatch, couple with electrostatic effects arising from polarization discontinuities at ferroelectric-dielectric interfaces. The coupled effects of mechanical and electrostatic boundary conditions modify the local symmetry of ferroelectric unit cells, generating stable self-organized and topologically nontrivial textures\cite{Junquera2023}, including polar skyrmions\cite{Das2019}, polar merons\cite{Wang2020}, and polar vortices\cite{Yadav2016}. These emergent polar textures host spatially varying unconventional dielectric properties\cite{Zubko2016,Yadav2019,Das2021}, and exhibit strong light-matter interactions\cite{Stoica2019,Gao2024} including sub-terahertz collective modes\cite{QianLi2021,HugoSubmitted}. The topologically nontrivial polar textures are also highly susceptible to electric fields\cite{Hong2018,Nahas2020,Das2021,Han2022,Zhu2022,Behera2022,Wang2024}. Under a moderate out-of-plane electric field, the polar skyrmions can be fully polarized to a single $c$-domain state\cite{Das2021,Zhu2022,Wang2024}, with tens of nanoseconds recovery time\cite{Das2021}. Despite extensive interest in these systems, little is known about their nonlinear light-matter interaction at ultrafast time scales. Potential large optical nonlinearities in polar textures remain a frontier awaiting exploration.  

Here, we focus on the polar-skyrmion phase in (PbTiO$_3$)$_n$/(SrTiO$_3$)$_n$ (PTO/STO) superlattices. The impact of mid-infrared laser excitation on the second-harmonic generation (SHG) process is explored. The mid-infrared excitation induces a giant transient SHG intensity from the polar skyrmions, exhibiting a fourth-power dependence on the excitation field. In addition, the SHG intensity oscillates at the mid-infrared driving frequency, with a dominant third-power dependence on the excitation field. This is attributed to an effective fourth-order optical response. The giant light-induced nonlinear polarization in the polar skyrmion is in stark contrast to its vanishing in-plane macroscopic polarization in equilibrium. A maximum modulation ratio of 350 is achieved with in-plane detections. This high contrast, sub-picosecond response time, and the absence of strict in-plane orientation constraints are crucial for potential applications such as photonic switches, terahertz-to-visible photon converters and terahertz detectors. More generally, we observe a strong higher-order optical nonlinearity across diverse polar textures in PTO/STO superlattices. These polar textures exhibit a strong and periodic variation of local permittivity, resemble metamaterials and photonic crystals. Investigating the large inherent higher-order optical nonlinearity in self-assembled polar textures could unlock a new paradigm for nonlinear optics\cite{Sun2025,Dogadov2022} and quantum photonic platforms\cite{Chang2014,Elshaari2020}, combining the scalability of bottom-up synthesis with topological-texture-enabled light-matter interactions.


\section*{Results}\label{sec2}
\begin{figure*}
\centering
\includegraphics[width=1\textwidth]{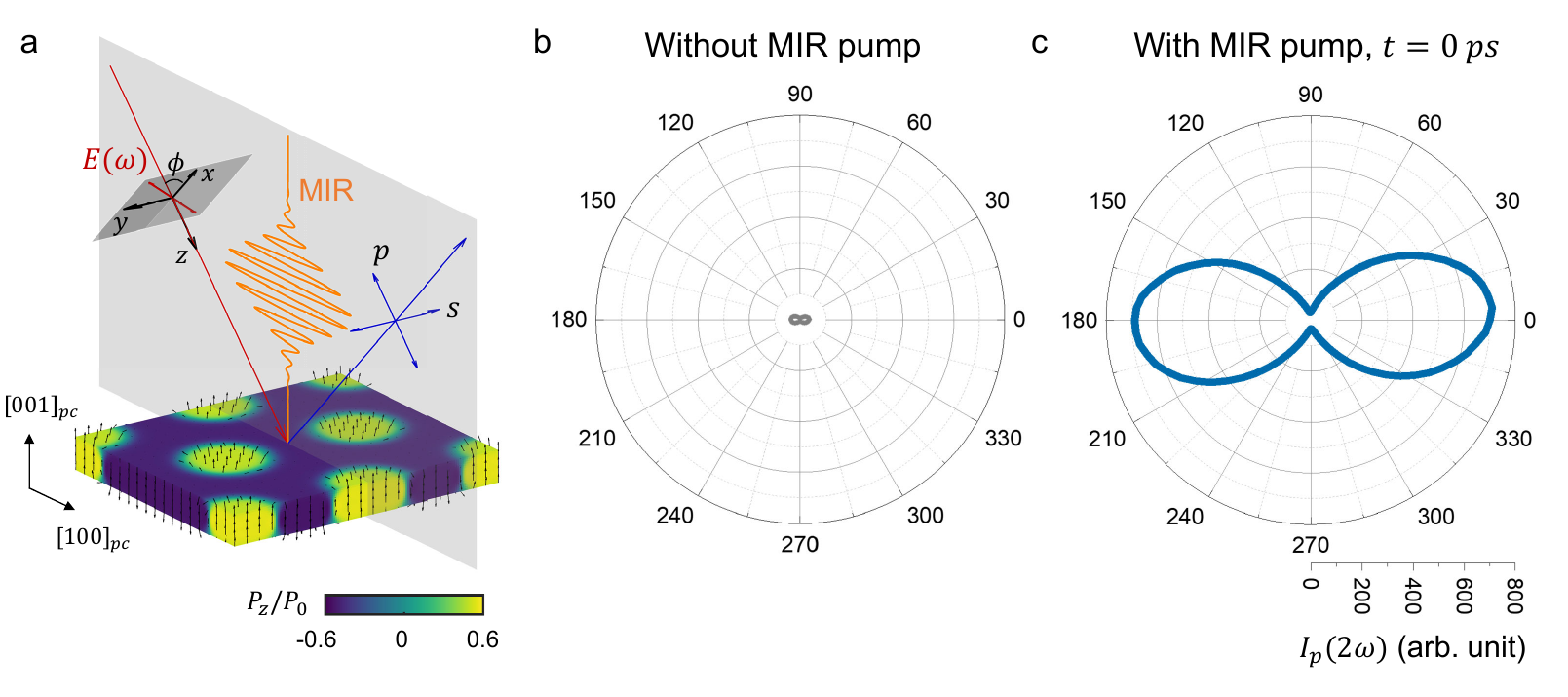}
\caption{\textbf{Strong light-induced SHG in the polar skyrmion.} \textbf{a,} In equilibrium, the polar skyrmion has a net ferroelectric polarization along the sample normal. We measured SHG polarimetry using a linearly polarized 800 nm fundamental beam with varying angle $\phi$ and two fixed detection polarizations ($p$ and $s$). Mid-infrared excitation (16.7 µm, 18 THz) is linearly polarized and comes at normal incidence. \textbf{b,} A weak but finite SHG signal is seen in equilibrium. \textbf{c,} With the mid-infrared excitation, the SHG intensity is enhanced by a factor of 20 near $\phi=0^\circ$ (along one cubic edge of the sample). Here $t=0$ ps is defined by the maximum response.}
\label{fig:GiantSHG}
\end{figure*}

\subsection*{Giant light-induced second-harmonic generation}\label{subsec2}
In the polar-skyrmion phase, the ferroelectric polarization vectors rotate smoothly from the center of a skyrmion to its boundary, forming a three-dimensional swirl pattern with a scale of $\sim$10 nm\cite{Das2019}. At the skyrmion core, the polarization vectors align parallel to each other and are normal to the sample surface. The spatial average over the entire skyrmion leads to a strongly suppressed in-plane macroscopic ferroelectric polarization. Room-temperature SHG polarimetry measurements were conducted on (PbTiO$_3$)$_{16}$/(SrTiO$_3$)$_{16}$ superlattices grown on SrTiO$_3$ (001) substrates known to possess polar skyrmions. SHG polarimetry was measured in reflection at 45$^\circ$ incidence along orthogonal directions (Fig.~\ref{fig:GiantSHG}a). The $s$ direction is parallel to the superlattice's surface. The $p$ direction is parallel to the incidence plane. Overall, the equilibrium SHG intensity is weak (Fig.~\ref{fig:GiantSHG}b and Supporting Information Fig. S2). The $p$ channel signal is slightly stronger, consistent with its sensitivity to the finite out-of-plane polar response. When driving the polar skyrmion out of equilibrium with mid-infrared (MIR) laser pulses (18 THz, 230 fs), a large SHG signal is observed (Fig.~\ref{fig:GiantSHG}c). The maximum transient SHG intensity is about 20 times greater than the equilibrium value. Further measurements show that the maximum transient SHG intensity is insensitive to both rotation of the sample and rotation of the linear MIR polarization. Rotating the MIR polarization by 90$^\circ$ brings the $s$ channel transient SHG intensity up to the same level as that in Fig.~\ref{fig:GiantSHG}c (Supporting Information Fig.~S5). Due to spatial cancellation of the winding in-plane polarization components, the equilibrium SHG intensity is an order of magnitude weaker in the $s$-channel than the $p$-channel (Fig. S2). This results in a maximum SHG intensity modulation ratio of 350 with parallel pump and probe polarizations.

\subsection*{Evidence of the higher-order nonlinearity}\label{subsec2}
\begin{figure*}[h!]
\centering
\includegraphics[width=0.85\textwidth]{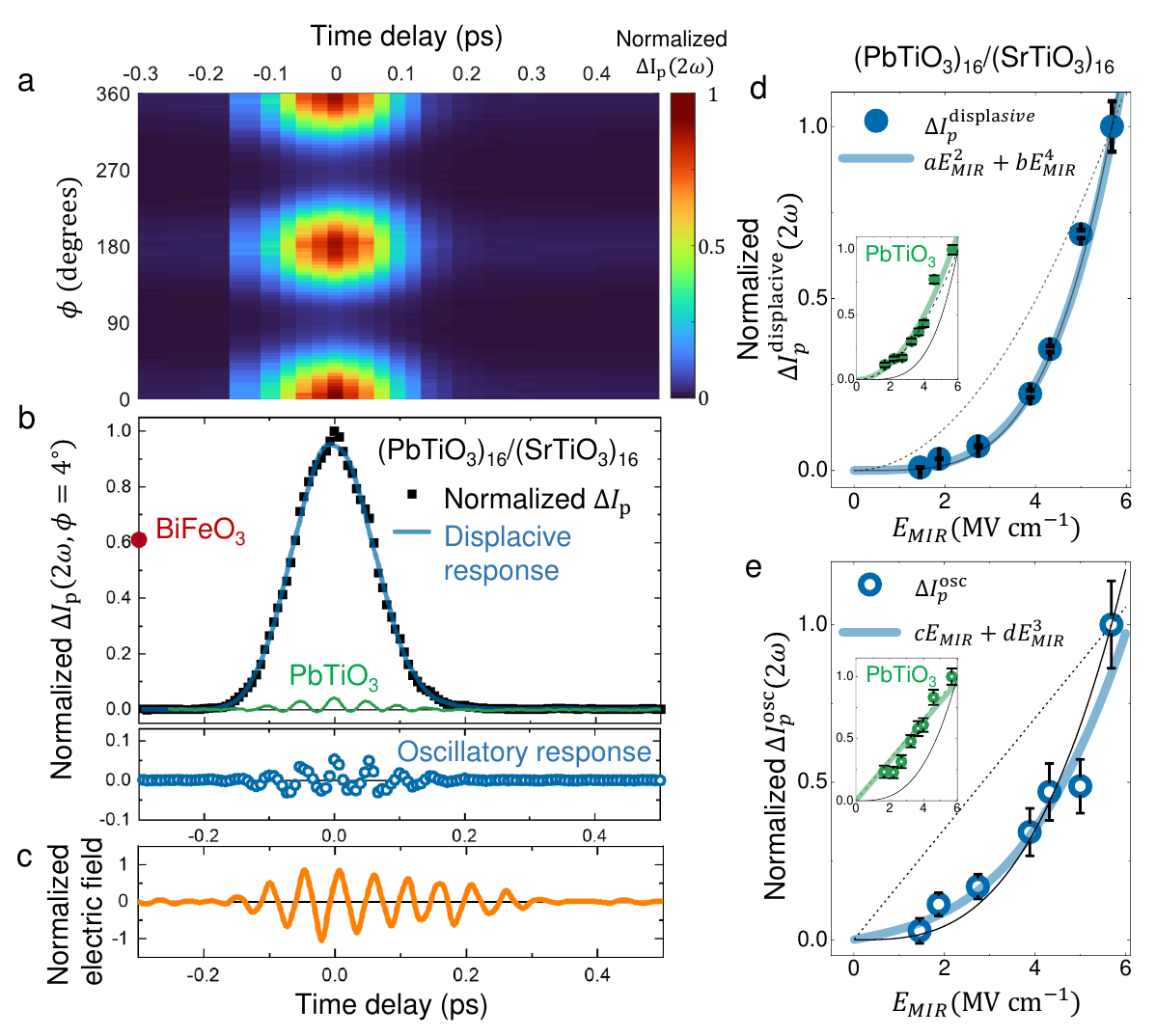}
\caption{\textbf{Evidence of the higher-order nonlinearity.} \textbf{a,} Normalized transient SHG polarimetry as a function of pump-probe time delay. \textbf{b,} Time evolution of $\Delta I_p(2\omega)$ at the maximum response angle. A weak oscillation can be resolved (blue open circles) after removing the displacive response (blue solid line). For comparison, we present the equilibrium SHG intensity from a [001]$_{pc}$ BiFeO$_3$, measured in the same setup without MIR excitation (red solid circle), together with the MIR-induced SHG intensity change of a 100-nm-thick PbTiO$_3$ film with ferroelectric polarization perpendicular to the film surface (green solid line). \textbf{c,} An electro-optical sampling characterization of the mid-infrared field. \textbf{d,} For the superlattice sample, the displacive response in $\Delta I_p(2\omega)$ shows a dominant quartic pump-field dependence. (Inset) In contrast, the PbTiO$_3$ film shows a quadratic pump-field dependence for the displacive response. \textbf{e,} For the superlattice, the oscillatory response follows a dominant cubic field dependence. (Inset) In contrast, the oscillatory response of the PbTiO$_3$ film scales linearly with the pump field. For both \textbf{d} and \textbf{e}, vertical error bars represent the standard deviation. The thin dashed/solid lines are guides for the eye on pure quadratic/quartic pump-field dependence for \textbf{d}, and pure linear/cubic pump-field dependence for \textbf{e}.} 
\label{fig:FluenceDep}
\end{figure*}

The transient SHG intensity is pronounced as compared with the equilibrium SHG from, for example, a single-domain BiFeO$_3$ crystal (Fig.~\ref{fig:FluenceDep}b and Fig.S3), a well-known ferroelectric with large polarization and large second-order optical nonlinearities. The transient SHG enhancement in the polar skyrmion sample has a lifetime comparable to the duration of the MIR pulse (Figs.~\ref{fig:FluenceDep}a to~\ref{fig:FluenceDep}c). In addition, the transient SHG intensity oscillates with time. The non-oscillatory component of the signal is fitted with a Gaussian function and defined as the displacive response. The oscillatory response is then obtained by subtracting the fitted displacive contribution from the raw data (Fig.~\ref{fig:FluenceDep}b, see Supporting Information S7 for details). The oscillatory response follows the MIR field (Fig.~\ref{fig:FluenceDep}c). A control experiment on a PbTiO$_3$ film with a net out-of-plane polarization shows a similar oscillatory response but a significantly weaker displacive response (Fig.~\ref{fig:FluenceDep}b and Fig. S4), highlighting the unique character of the superlattices. As the MIR excitation used in this work is far away from the sub-terahertz collective modes\cite{LiarXiv,HugoSubmitted}, no other low-frequency oscillations other than that at the MIR driving frequency are observed. 

We next examined the pump-field dependence of the transient SHG response. When using a polynomial fit, the transient SHG intensity scales predominantly to the fourth-power of the MIR peak field (Fig.~\ref{fig:FluenceDep}d). The oscillatory signal shows a dominant third-order field dependence (Fig.~\ref{fig:FluenceDep}e). These results indicate that higher-order nonlinear optical processes are present. Under the same MIR excitation, the displacive (Fig.~\ref{fig:FluenceDep}d inset) and oscillatory (Fig.~\ref{fig:FluenceDep}e inset) responses of the PbTiO$_3$ film with $c$ domain polarization exhibit the quadratic and linear dependencies on the pump field, respectively.

\begin{figure*} [h]
\centering
\includegraphics[width=1\textwidth]{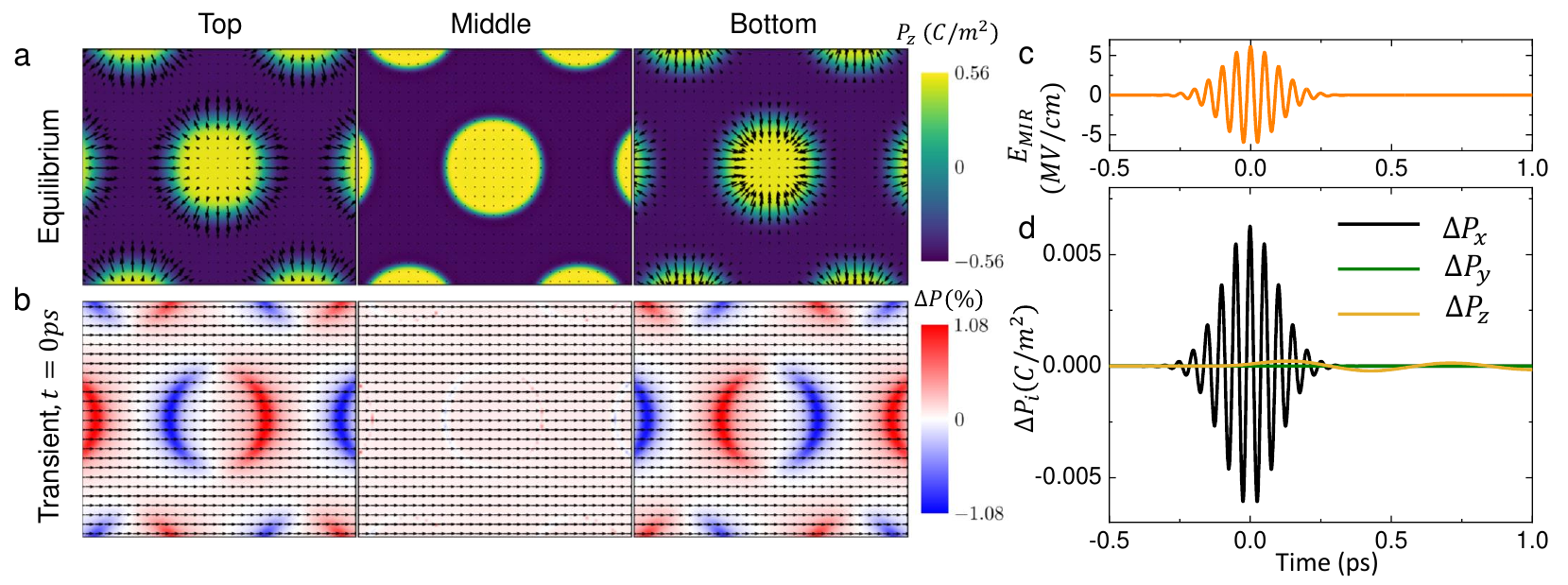}
\caption{\textbf{Phase-field simulation of the polar-skyrmion phase.} \textbf{a,} Equilibrium ferroelectric polarization distribution for the top, middle and bottom of the PTO layer. \textbf{b,} Light-induced change in ferroelectric polarization at maximum response ($t=0ps$). At this instant, the pump electric field points to the right. The black arrows indicate the pump-induced change in the in-plane ferroelectric polarization. The color scale shows the percentage change in ferroelectric polarization magnitude, which is largest at the top and bottom surfaces. \textbf{c,} The excitation pulse used in the simulation. \textbf{d,} The average light-induced ferroelectric polarization change along $x$ (pump polarization direction), $y$ and $z$ directions.} 
\label{fig:Simulation}
\end{figure*}

We used phase-field simulations to investigate the polarization dynamics in the light-driven state for the polar skyrmions. Figure~\ref{fig:Simulation}a shows the stabilized distribution of ferroelectric polarization at the top, middle, and bottom of the PTO layers, exhibiting polar-skyrmion structure. The maximum light-induced change in the ferroelectric polarization is shown in Fig.~\ref{fig:Simulation}b. At time zero, the electric field of the excitation pulse reaches a maximum and is along the $+x$ direction. From Fig.~\ref{fig:Simulation}d it is seen that the entire PTO layer develops a transient in-plane ferroelectric polarization following the MIR field (Fig.~\ref{fig:Simulation}c). The transient change in the magnitude of ferroelectric polarization is strongest at the top and bottom surface of the PTO layers and along the skyrmion walls. No sizable quasi-static change in the spatially averaged ferroelectric polarization is observed in the simulation (Fig.~\ref{fig:Simulation}d). Overall, the average ferroelectric polarization change exhibits a stronger oscillation along the pump-field direction (Fig.~\ref{fig:Simulation}d), while a much weaker and slower oscillation is seen for the ferroelectric polarization change along $z$. The magnitude of the ferroelectric polarization change is small with respect to the maximum local electric dipole strength (Fig.~\ref{fig:Simulation}a). The rapid polarization oscillation driven by the MIR field is consistent with the oscillatory component observed in the transient SHG signal (Fig.~\ref{fig:FluenceDep}b).

\subsection*{Polarization dynamics in various polar textures}\label{subsec3}
\begin{figure*}
\centering
\includegraphics[width=0.9\textwidth]{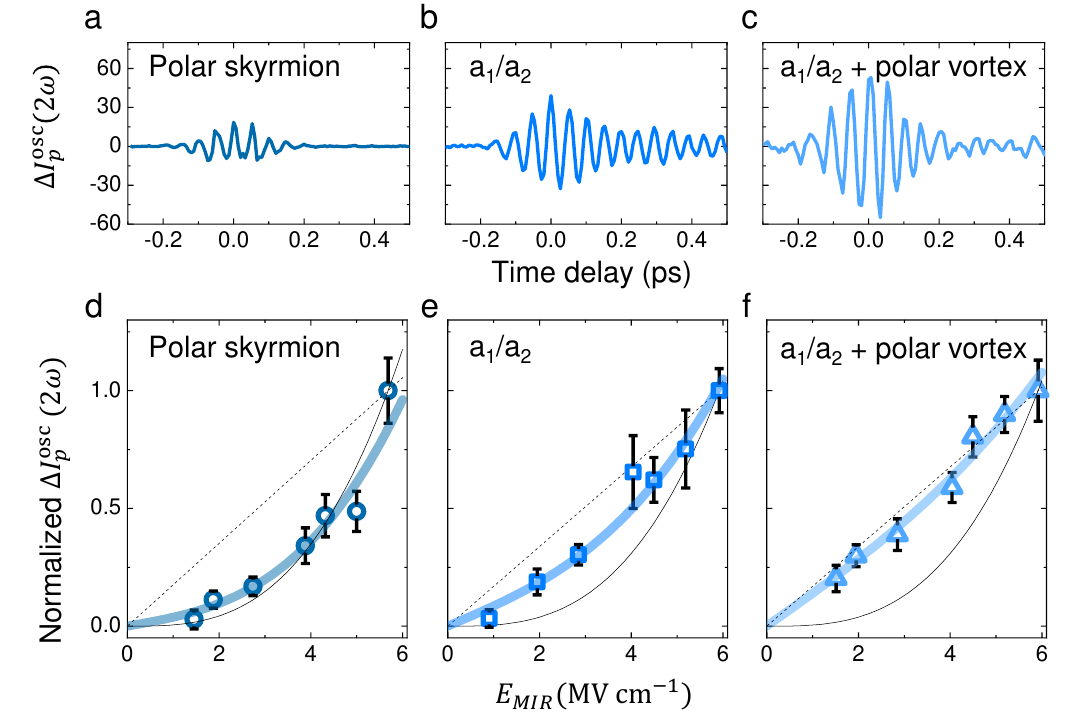}
\caption{\textbf{Signature of the higher-order optical response in various polar phases.} The top panels (\textbf{a-c}) show the time-domain response at the maximum pump fluence. The bottom panels (\textbf{d-f}) show the pump-field dependence of the oscillation intensity. The thick solid lines are fits with a mixed linear--cubic dependence (see Supporting Information Figure S9 and Table S3). The thin dashed/solid lines are guides for the eye in a pure linear/cubic pump-field dependence respectively.} 
\label{fig:SampleDep}
\end{figure*}

In addition to the polar skyrmions, other polar phases were examined including a mixed classical ferroelectric ($a_1$/$a_2$) and polar-vortex phase\cite{Damodaran2017}, and a pure $a_1$/$a_2$ phase\cite{a1a2}. All three polar phases exhibit strong light-induced SHG (Supporting Information S5) with oscillations at the MIR excitation frequency. Because nonlinear-phononic contributions may also occur in phases with a well-defined macroscopic polarization\cite{a1a2}, we focus here on the oscillatory response and its pump-field dependence when comparing the higher-order optical response across phases. The oscillation intensity at the maximum pump fluence is shown in Figs.~\ref{fig:SampleDep}a-\ref{fig:SampleDep}c. The strongest oscillation is seen in the mixed $a_1$/$a_2$ and polar-vortex phase. The corresponding pump-field dependence of the oscillation intensity is shown in Figs.~\ref{fig:SampleDep}d-\ref{fig:SampleDep}f. Cubic pump-field dependence dominates the polar-skyrmion phase, weakens in pure $a_1$/$a_2$, while a nearly linear dependence is seen in the mixed phase.


\begin{figure*} [h]
\centering
\includegraphics[width=1\textwidth]{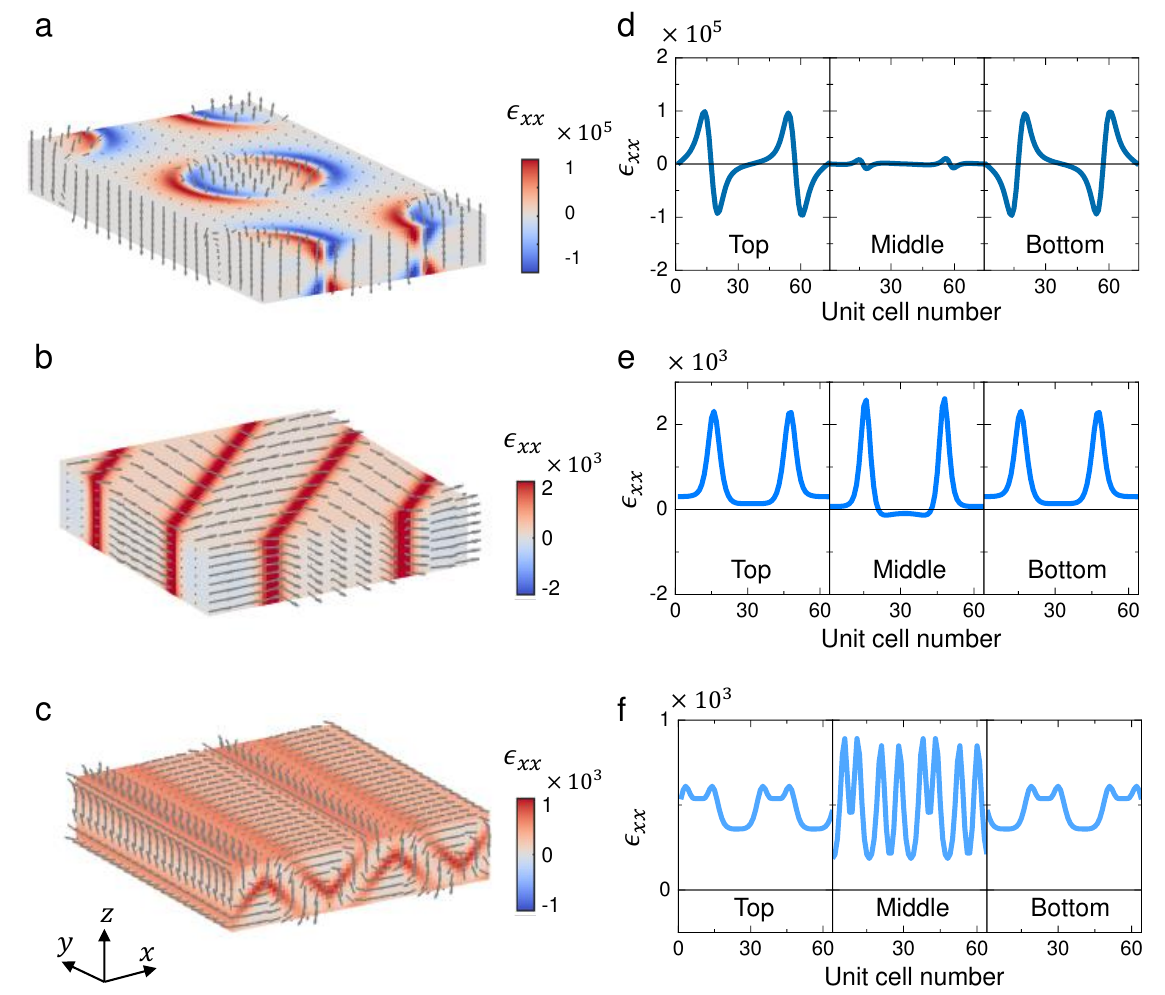}
\caption{\textbf{Phase-field simulations of the local static permittivity distribution.} Ferroelectric polarization distribution (arrows) and the local static permittivity $\epsilon_{xx}$ (color coded) of the PTO layer for the polar-skyrmion phase (\textbf{a}), the $a_1/a_2$ phase (\textbf{b}), and the polar-vortex phase (\textbf{c}). Spatial variation of $\epsilon_{xx}$ along the $x$ direction and through the center of the corresponding 3D plots for the top, middle and bottom of the PTO layer for the polar-skyrmion phase (\textbf{d}), the $a_1/a_2$ phase (\textbf{e}), and the vortex phase (\textbf{f}).} 
\label{fig:SimulationESP}
\end{figure*}

We used phase-field simulations to model the ferroelectric polarization and corresponding in-plane static permittivity ($\epsilon_{xx}$) of the polar-skyrmion, $a_1/a_2$, and polar-vortex phases. All three polar phases exhibit a strong spatial variation in $\epsilon_{xx}$. Alternating regions of large negative and positive values of $\epsilon_{xx}$ are observed near the domain walls for polar skyrmions (Fig.~\ref{fig:SimulationESP}a). In addition, $\epsilon_{xx}$ varies strongly along the film thickness direction, as evidenced by the side profile of the 3D plot in Fig.~\ref{fig:SimulationESP}a and the line-cuts in Fig.~\ref{fig:SimulationESP}d. In contrast, regions of positive local $\epsilon_{xx}$ are seen in the $a_1/a_2$ phase along the domain walls (Fig.~\ref{fig:SimulationESP}b) with small variations along the film thickness (Fig.~\ref{fig:SimulationESP}e). Positive local $\epsilon_{xx}$ is also seen for the polar-vortex phase (Figs.~\ref{fig:SimulationESP}c and \ref{fig:SimulationESP}f). Here we simulated a pure vortex phase for simplicity. We anticipate that additional dielectric heterogeneity may arise at the domain boundaries in the $a_1/a_2$ and polar-vortex mixed phase\cite{Damodaran2017}, producing a broader spatial variation of $\epsilon_{xx}$ than in the pure $a_1/a_2$ and pure polar-vortex phases. Overall, the local permittivity varies strongly and periodically throughout the PTO layer in all three polar phases, creating a natural analog to engineered metamaterials and photonic crystals.

\section*{Discussion}\label{sec3}
The key findings of this work are the giant ultrafast second-harmonic generation from the polar skyrmions, and its unconventional pump-field dependence: a predominant quartic-field dependence for displacive response, and a cubic-field dependence for the oscillatory response. We now consider possible underlying mechanisms. For simplicity, throughout the discussion section the polarization $P$ and the electric field $E$ are taken to be scalar quantities\cite{BoydBook}. The tensor contraction form can be found in the Supporting Information Section S3. We first consider nonlinear-phononic rectification\cite{Disa2021}. If a driven infrared-active phonon couples quadratically to another structural or polar coordinate, the induced displacement can scale as the square of the pump field. A corresponding quadratic change of the SHG field can generate both quadratic and quartic contributions to the measured SHG intensity through interference with the equilibrium SHG field. Pump-field scaling alone therefore does not uniquely distinguish this mechanism. Although the nonlinear phononics effect in PbTiO$_3$ is predicted to be prominent\cite{Subedi2015}, the lifetime of the transient state is equal to or longer than the lifetime of the driven infrared-active phonon (typically in the order of a few picoseconds).\cite{Mankowsky2017}. In contrast, the displacive response observed here follows the $\sim230$~fs MIR excitation and decays within less than 1~ps (Figs.~\ref{fig:FluenceDep}b and \ref{fig:FluenceDep}c). We therefore conclude that nonlinear-phononic rectification is unlikely to be the dominant origin of the observed ultrafast response.

In turn, we investigate nonlinear optical processes that directly match the excitation pulse duration. We start from the process which involves the third-order susceptibility $\chi^{(3)}$. A short-lived SHG with oscillations at the pump frequency may arise from the electric-field-induced SHG (EFISH) process\cite{Grishunin2017}. For materials with a nonzero second-order susceptibility, $\chi^{(2)}$, the SHG intensity is given by
\begin{equation}\label{eq:chi2chi3}
I\propto\Bigl\{\left[\chi^{(2)}+\chi^{(3)}E(\omega_1)\right]E(\omega)E(\omega)\Bigr\}^2.
\end{equation}
Here, $\omega_1$ is the frequency of the pump field, while $\omega $ is the fundamental frequency of the optical probe beam. This equation has three terms (see Supporting Information Section S7): (1) A term representing the SHG intensity at equilibrium. (2) A homodyned term containing $\left[\chi^{(3)}E(\omega_1)\right]^2$, which includes a displacive term and an oscillatory term (the SHG intensity oscillates at $2\omega_1$, filtered out in our detection scheme). Both intensities scale quadratically with the pump-field amplitude. (3) A heterodyned term proportional to $\chi^{(2)}\chi^{(3)}E(\omega_1)$, which oscillates at $\omega_1$ with an intensity that scales linearly with the pump-field amplitude. The control experiment on a PbTiO$_3$ film exhibits precisely this scaling behavior (Fig.~\ref{fig:FluenceDep}d and \ref{fig:FluenceDep}e insets). In the skyrmion case, the transient SHG intensity oscillation at the pump frequency follows a leading cubic pump-field dependence (Fig.~\ref{fig:FluenceDep}e), and the displacive response shows a predominant quartic pump-field dependence (Fig.~\ref{fig:FluenceDep}d). Therefore, higher-order nonlinear process needs to be considered.

Continuing on, the process involving the fourth-order optical susceptibility $\chi^{(4)}$ is examined. In this case, the second-harmonic generation intensity is given by
\begin{equation}\label{eq:chi2chi3chi4}
I\propto\Bigl\{\left[\chi^{(2)}+\chi^{(3)}E(\omega_1)+\chi^{(4)}E(\omega_1)E(\omega_1)\right]E(\omega)E(\omega)\Bigr\}^2.
\end{equation}
which contains six terms (see Supporting Information Section S7). Among these, three terms contribute to the displacive response: one homodyned term with $\left[\chi^{(3)}E(\omega_1)\right]^2$ and one heterodyned term with $\chi^{(2)}\chi^{(4)}E(\omega_1)E(\omega_1)$ which follows a quadratic pump-field dependence, and one homodyned term containing $\left[\chi^{(4)}E(\omega_1)E(\omega_1)\right]^2$ which follows a quartic pump-field dependence. This agrees with observations as shown in Fig.~\ref{fig:FluenceDep}d. Two heterodyned terms create oscillations at the pump frequency: the term containing $\chi^{(2)}\chi^{(3)}E(\omega_1)$ exhibits a linear pump-field dependence, and the term containing $\chi^{(3)}\chi^{(4)}E(\omega_1)E(\omega_1)E(\omega_1)$ follows a cubic pump-field dependence. This agrees with observation as shown in Fig.~\ref{fig:FluenceDep}e. The above mixed processes also produce oscillations at $n\omega_1$ ($n=2,3,4$), which are beyond the detection bandwidth in the measurement.

At the microscopic level, the term containing $\chi^{(4)}$ in Eq.~\ref{eq:chi2chi3chi4} may arise from a cascaded process involving lower-order nonlinearities. For materials with a nonzero $\chi^{(2)}$, an optical pump field $E(\omega_1)$ can lead to transient polarization through the mechanism of optical rectification\cite{Zhang2024}, described by 
\begin{equation}\label{eq:OR}
P_{\mathrm{OR}}(0)\propto\chi^{(2)}E(\omega_1)E(-\omega_1).
\end{equation}
The associated rectified electric field, $E_{\mathrm{OR}}(0)$, can subsequently enter a third-order interaction,
\begin{equation}\label{eq:chi3andOR}
P(2\omega)\propto\chi^{(3)}E(\omega)E(\omega)E_{\mathrm{OR}}(0) .
\end{equation}
Because $E_{\mathrm{OR}}(0)\propto E^2(\omega_1)$, this cascaded pathway has the same pump-field dependence as a direct fourth-order process. To further investigate a potential optical rectification effect, we conducted terahertz (THz) emission measurements under MIR excitation, which probes the radiative optical-rectification response. No sample-generated far-field THz emission was resolved under our measurement conditions, and the sample exhibited negligible THz absorption (Fig.~S7). The THz-emission measurement constrains a radiative macroscopic rectification component in the measured geometry but does not exclude local or predominantly out-of-plane rectification. Within the effective \textit{4mm} description, the strong spatial cancellation of the winding in-plane response coexists with a finite out-of-plane polar component in the skyrmion phase. An in-plane MIR field then generates a rectified polarization along the sample normal $z$ (Supporting Information Section S6), consistent with the absence of THz emission probed along the MIR propagation direction. Thus, both direct fourth-order and cascaded contributions can be the underlying mechanisms for our observation.

More generally, the transient SHG signal is the far-field result of a coherent superposition of local nonlinear polarization sources across the textured medium. Accordingly, the $\chi^{(3)}$ and $\chi^{(4)}$ terms discussed here should be regarded as effective susceptibilities, consistent with the concept of an effective nonlinear optical susceptibility commonly used in photonic crystals and metamaterials. The third- and fourth-order optical responses therefore reflect both the material properties of the polar texture and the strong-field mixing conditions. As shown in Supporting Information Fig.~S9d--S9f and Table S3, the strong MIR field is important for making the effective fourth-order contribution visible. Nevertheless, the PbTiO$_3$ control film under the same MIR excitation is dominated by the lower-order field dependence, showing that field strength alone does not account for the superlattice behavior. The fits further show that the degree of deviation from a purely linear oscillatory response varies across the polar phases (Fig.~S9d--S9f and Table S3), consistent with a texture-dependent nonlinear response. Although the skyrmion phase has the weakest absolute oscillatory signal among the polar phases (Figs.~\ref{fig:SampleDep}a-\ref{fig:SampleDep}c), its equilibrium SHG intensity is about 10--20 times smaller (Fig.~S6), resulting in the strongest normalized oscillation intensity (Figs.~S9a--S9c).

We finally consider material properties that may contribute to the enhanced effective nonlinear response. Previous studies report strong spatial variations of the static permittivity in polar skyrmions, including negative values near the skyrmion walls, together with a high susceptibility to external electric fields\cite{Das2021,Wang2024}. These observations primarily concern the out-of-plane dielectric response, whereas the MIR excitation used here is polarized within the plane of the skyrmion-containing layers. Negative local static permittivity has been linked to a flat Landau free-energy landscape\cite{Yadav2019,Das2021}. Although the MIR pump lies near a PTO optical-phonon frequency\cite{Freire1988}, the sub-picosecond response discussed above disfavors a long-lived phonon-driven displaced state\cite{a1a2}. We notice that all three polar phases exhibit strong and periodic spatial variations of the calculated static permittivity. Among three phases, the polar-skyrmion phase has alternating regions of large negative and large positive permittivity, which may lead to the largest permittivity renormalization. Although the impact of polar topology on the dielectric dispersion is beyond the scope of this research, a largely suppressed static permittivity could reshape the dielectric dispersion and fulfill the epsilon-near-zero condition, which enhances the effective field within the material. For example, in metamaterials and photonic crystals, epsilon-near-zero conditions are achieved by mechanically engineering periodic nanostructures with a designed spatial variation of permittivity. This is the key mechanism that leads to dramatically enhanced nonlinear optical response\cite{Luk2015,Liberal2017,Reshef2019,Kinsey2019}. The advantage here is that the spatial variation of negative and positive permittivity regions is inherent to the polar texture. As the negative permittivity is seen for various polar phases, depending on applied field direction\cite{Yadav2019, Das2021}, polar textures could potentially serve as next-generation nanophotonic materials, where large optical nonlinearities emerge from self-assembled polar topologies. Establishing a quantitative connection between the calculated static permittivity patterns and the observed nonlinear enhancement will require frequency-dependent dielectric measurements or calculations together with a local-field analysis. We therefore regard the nanoscale dielectric heterogeneity as a possible materials-level contributor to the effective nonlinear response rather than as an identified microscopic mechanism. Clarifying this connection may provide a route to using the self-organized dielectric landscape of polar textures to engineer nonlinear optical responses without externally patterned photonic structures.  

\section*{Conclusions}\label{sec4}

To conclude, this work has demonstrated pronounced nonlinear polarization in the polar skyrmions. Ambient temperature operation, ultrafast response time, and the absence of in-plane orientation constraints for the polar skyrmions are significant benefits as mid-infrared sensors. In addition, the large contrast in orthogonal detection channels allows polarization-selective sensing. Furthermore, harnessing the higher-order nonlinearity may unlock photonics applications that leverage multiphoton interactions and non-degenerate wave mixing processes such as terahertz-to-visible photon converters. More generally, a strong higher-order optical nonlinearity is observed in various polar textures, including the $a_1/a_2$ and the mixed polar vortex phases. All polar textures exhibit a periodic modulation of permittivity, resembling nano-engineered metamaterials and photonic crystals. The inherent strong nonlinearity from the rich polar topologies could uniquely mediate nonlinear light-matter interactions, which positions polar textures as novel nanophotonic materials for quantum information processing and high-speed optical communications.

\section*{Methods}\label{sec5}
\subsection*{Sample specifications}\label{subsec5}
The polar skyrmion sample is a 100-nm-thick [(PbTiO$_3$)$_{16}$/(SrTiO$_3$)$_{16}$]$_8$ superlattice grown on SrTiO$_3$ (001) substrate. The $a_1/a_2$ phase sample is a [(PbTiO$_3$)$_{16}$/(SrTiO$_3$)$_{16}$]$_8$ superlattice on TbScO$_3$ (110) substrate. The mixed phase sample is grown on  DyScO$_3$ (110) substrate with a 3-unit-cell SrRuO$_3$ buffer layer followed by [(PbTiO$_3$)$_{16}$/(SrTiO$_3$)$_{16}$]$_7$(PbTiO$_3$)$_{16}$(SrTiO$_3$)$_6$. Samples were fabricated using pulsed laser deposition with a KrF excimer laser as described previously in Refs.\cite{Yadav2016,Das2019}. Films were deposited from Pb$_{1.2}$TiO$_3$ and SrTiO$_3$ targets at ~700$^\circ C$ in 100 mTorr O$_2$ pressure with a fluence of ~2 J cm$^{-2}$ on (110)-oriented DyScO$_3$, (110)-oriented TbScO$_3$, and (001)-oriented SrTiO$_3$ substrates for the mixed phase, $a_1/a_2$, and skyrmion samples respectively. Layer thicknesses were continuously monitored \textit{in situ} using reflection high-energy electron diffraction (RHEED) along the [100] direction and confirmed \textit{ex situ} using lab-based x-ray diffraction (Supporting Information section S1) or synchrotron-based three-dimensional-reciprocal space mapping as reported in Refs.\cite{Das2019,Stoica2019,HugoSubmitted}. The PbTiO$_3$ film used as the control sample was grown on SrTiO$_3$ (001) substrates under the same growth conditions used for the superlattices. The film thickness was monitored \textit{in situ} by RHEED during deposition. The PbTiO$_3$ film exhibits a predominantly out-of-plane polarization.

The BiFeO$_3$ is predominantly a single domain single crystal, cut along the pseudocubic [001] direction. The details of the growth and sample characterization can be found in Ref.\cite{DanielBFO}.

\subsection*{MIR pump-SHG probe setup}\label{subsec5}
Room-temperature SHG measurements were performed in reflection geometry with 45$^\circ$ incidence. The fundamental beam for SHG has a center wavelength of 800 nm, with 6~nJ energy and 40~fs pulse duration. A half-wave plate was used to rotate the polarization of the fundamental beam. The probe spot size was 78 micron in diameter on the sample. The SHG beam was collimated at 40 mm after the sample to reduce spatial separation of $\chi^{(2)}$ and higher-order nonlinear processes. A Glan-Taylor prism was used to select the desired polarization for detection in a photo-multiplier tube. The SHG bandwidth is 10 nm (full width at half maximum, FWHM), as measured by the SHG process in a BBO crystal. A 400~nm bandpass filter with a 10~nm FWHM was used for SHG detection. The photomultiplier records the SHG intensity integrated over the spectral window transmitted by the filter. Due to the finite SHG bandwidth, contributions from higher-order nonlinear processes may enter the detected signal depending on their spectral overlap with the filter transmission. The relevant frequency components are summarized in Supporting Information Table S2.

MIR pump pulses were generated by optical parametric amplification and subsequent difference frequency generation in a GaSe crystal. The MIR spot size on sample was 176 micron in diameter, with a maximum 11~mJ~cm$^{-2}$ fluence at the longest wavelength (16.7 micron/18 THz). The MIR pump was at normal incidence on the sample.

\subsection*{Phase-field simulations}\label{subsec6}
Phase-field simulations\cite{Akamatsu2018,Yang2020} were conducted to investigate the formation of the polar-skyrmion phase, the $a_1/a_2$ phase and the polar-vortex phase in ferroelectric superlattices. The system was modeled using a three-dimensional mesh ($289 \times 250 \times 32$ for the polar-skyrmion phase, $64 \times 64 \times 32$ for the $a_1/a_2$ phase, and $128 \times 128 \times 32$ for the polar-vortex phase) with a grid spacing of 0.4 nm. The polarization components $P_i$ (where $i \in \{1,2,3\}$) were used as the order parameters and evolved according to the polarization dynamic equation:
\begin{equation}
\mu \dfrac{\partial^2 P_i}{\partial t^2} + \gamma \dfrac{\partial P_i}{\partial t} = -\dfrac{\delta F}{\delta P_i}
\end{equation}
where $\mu$ stands for the effective inertial coefficient of polarization, $\gamma$ is the damping constant, and $F = \int f dV$ represents the total free energy over the simulated volume.  The free energy includes contributions from Landau energy, electrostatic energy, elastic energy, and gradient energy:
\begin{equation}
    F = \int\left( f_\mathrm{Landau}+f_\mathrm{elastic}+f_\mathrm{electric}+f_\mathrm{gradient}\right) dV.
\end{equation}

Further details on the energy functions, numerical implementation, and simulation parameters can be found in the Supporting Information. Periodic boundary conditions were applied in all three spatial directions. The film was composed of alternating 16-unit-cell layers of PbTiO$_3$ and SrTiO$_3$ by changing the material coefficients using a sharp interface description. Each layer of the superlattice is coherently strained to the SrTiO$_3$ substrate which subjects the SrTiO$_3$ layer to $0\%$ strain, and the PbTiO$_3$ layer to $-1.34\%$ strain for the polar-skyrmion phase, and no strain for the $a_1/a_2$ phase. For the polar-vortex phase, the strain is anisotropic. The strain components are $\epsilon_{xx} = 1.2\%$ and $\epsilon_{yy} = 1.05\%$ for SrTiO$_3$, and $\epsilon_{xx} = -0.12\%$ and $\epsilon_{yy} = -0.275\%$ for PbTiO$_3$. The local static permittivity $\epsilon_{ii}$ ($i=x,y,z$) is evaluated by applying a small $dc$ electric field, and using $\epsilon_0\epsilon_{ii}=K_{ij}+\frac{\Delta P_i}{\Delta E_i}$, in which $\epsilon_0$ is the permittivity of free space, $K_{ij}$ is the dielectric permittivity tensor components, $\Delta P_i$ and $\Delta E_i$ are the local ferroelectric polarization change and local electric field change, respectively\cite{Das2021}. Because this quantity is evaluated using a small dc electric-field perturbation, it characterizes the static dielectric response of the phase-field model and should not be interpreted as the frequency-dependent complex permittivity at the optical frequencies used in the experiment.

To study the light-induced dynamics of skyrmions, an ultrashort laser pulse with a peak electric field $E_0 =6.14$ MV/cm and a Gaussian temporal profile (FWHM = $230$ fs) was applied. The light–matter interaction was introduced as a time-dependent uniform electric field. Initial polarization conditions included random noise of magnitude 0.0001 C/m$^2$ to mimic finite-temperature fluctuations during annealing.

\section*{Supporting Information}
The Supporting Information is available online.\\
Sample characterization; Equilibrium and pump-induced SHG; SHG polarimetry fit; Pump polarization dependence; Polar texture dependence; Absence of THz emission; Heterodyned and homodyned SHG components; Pump wavelength dependence; Phase-field simulation.
\section*{Acknowledgments}
We acknowledge Andrea Cavalleri, Aaron Lindenberg, Alexander von Hoegen and Albert Liu for discussions. D.A.B.L., D.S.G. and W.H. acknowledge support from the U.S. Department of Energy, Office of Basic Energy Sciences Early Career Research Program under Award Number DE-SC0021305. D.S.G. and W. H. additionally acknowledge support from the U.S. Department of Energy, Basic Energy Science, grant DE-SC0026026. X.L., V.A.S., L.W.M., L-Q.C. and V.G. acknowledge support from the U.S. Department of Energy, Basic Energy Science, grant DE-SC0012375. A.S., L.W.M. and R.R. additionally acknowledge the support of the Army Research Office under the ETHOS MURI via cooperative agreement W911NF-21-2-0162. A.R. acknowledges the support of the NSF Graduate Research Fellowship Program under Grant No. DGE1255832. S.D. acknowledges the Scheme for Transformational and Advanced Research in Sciences (MoESTARS/STARS-2/2023-0048), DST- FIST (SR/FST/ET-II/2023/1233 (C), and the Indian Institute of Science start up grant for financial support. M.T. would
like to acknowledge IISc doctoral Fellowship.
This work was supported by the W. M. Keck foundation grant to the Keck Center for Quantum Magnetism at Rutgers University. The computational work reported on in this paper was performed on the Shared Computing Cluster which is administered by Boston University's Research Computing Services.

\section*{Author contributions}
W.H. conceived the project, designed and performed the pump-probe measurements. D.A.B.L. fit the SHG polarimetry data with support from X.L. and V.G.. D.A.B.L. and D.S.G. analyzed the SHG fluence dependence. D.A.B.L. performed the phase-field simulations, with support from A.R. and L.-Q.C.. X.X. and S-W.C. grew the BiFeO$_3$ sample. M.T. and S.D. synthesized the PbTiO$_3$ film. S.D., P.M., A.S., L.M. and R.R. synthesized the PTO/STO superlattices. S.D., P.M., and A.S. conducted lab-based x-ray characterizations. V.A.S. and V.G. conducted synchrotron-based three-dimensional-reciprocal space mapping. W.H. drafted the main manuscript, and together with D.A.B.L. drafted the Supporting Information; all authors contributed to reviewing and editing the final paper.

\section*{Competing interests}
The authors declare that they have no competing interests.

\section*{Data Availability}
All data needed to evaluate the conclusions in the paper are present in the paper and/or the Supporting Information. 


\clearpage

\clearpage 
\begin{figure}[p]
    \centering
    \addcontentsline{lof}{figure}{For Table of Contents Only}
    
    \includegraphics[width=0.8\textwidth]{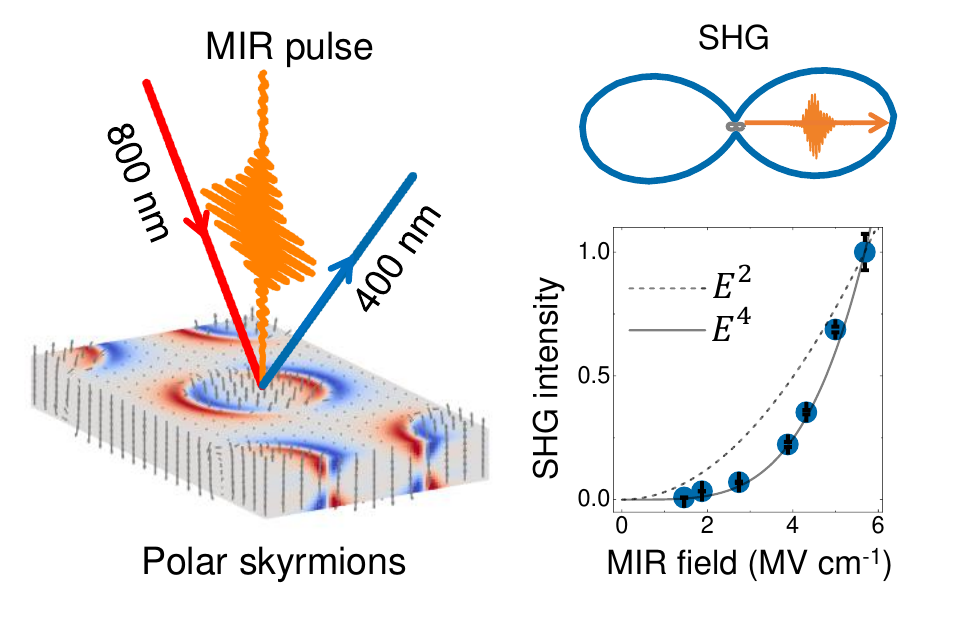}
    
    \caption*{For Table of Contents Only} 
\end{figure}

\clearpage 

\setcounter{section}{0} 
\renewcommand{\thesection}{S\arabic{section}}

\setcounter{figure}{0}
\setcounter{table}{0}
\renewcommand{\thefigure}{S\arabic{figure}}
\renewcommand{\thetable}{S\arabic{table}}

\section*{Supplementary Material}

\input{SI} 

\bibliography{sn-bibliography}
\end{document}

%% file: SI.tex
\tableofcontents
\clearpage
\markboth{}{}

\newpage
\section{Sample characterization}\label{sec1}

\begin{figure*}[h!]
\centering
\includegraphics[width=1\textwidth]{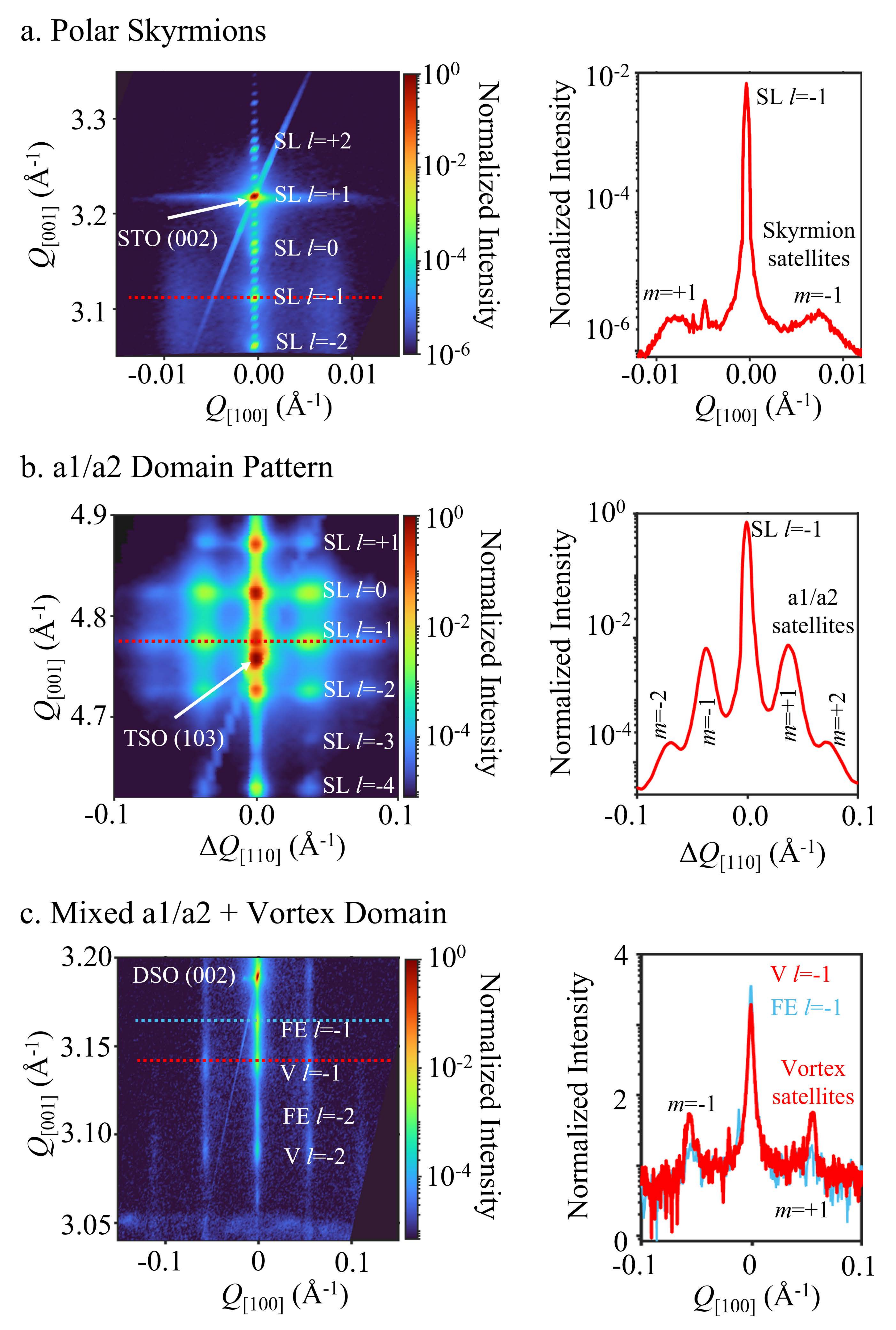}
\caption{\textbf{X-ray characterization of three superlattice samples.} } \label{fig:Xray}
\end{figure*}

Superlattice (SL) samples used in this study were characterized by either lab-based x-ray diffraction or synchrotron-based three-dimensional-reciprocal space mapping, as shown in Figure \ref{fig:Xray}. 

For the polar skyrmion sample, the left panel of Fig.~\ref{fig:Xray}a is the reciprocal space map (RSM) in the $Q_{[100]}–Q_{[001]}$ plane around the SrTiO$_3$ (STO) (002) reflection. The RSM shows the STO (002) reflection, along with several orders of (PbTiO$_3$)$_n$/(SrTiO$_3$)$_n$ SL reflections, labeled as $\pm l$. The superlattice peaks are separated by $\Delta Q_{[001]}=2\pi/d$. Here $d$ is the thickness of the total repeating units. The superlattice peaks are labeled such that the SL (002) reflection is $l$=0. Positive and negative integers are assigned to the peaks at higher and lower $Q_{[001]}$, respectively. The right panel of Fig.~\ref{fig:Xray}a is a line cut along Q$_{[100]}$ direction across SL $l=-1$ reflection (red dashed line in Fig.~\ref{fig:Xray}a left panel). The two broad satellite peaks around the SL $l=-1$ reflections are the domain diffuse scattering from the skyrmion periodicity. 

For the $a_1/a_2$ sample, synchrotron-based RSM is shown in Fig.~\ref{fig:Xray}b. The left panel is the RSM in the $Q_{[001]}–\Delta Q_{[110]}$ plane around the TbScO$_3$ (TSO) (103) reflection. Here $\Delta Q_{[110]}$ is defined as $Q_{[110]}-Q_{[110]}$(TSO (002)). The RSM contains the TSO (103) reflection, along with multiple orders of (PbTiO$_3$)$_n$/(SrTiO$_3$)$_n$ SL reflections, labeled as $\pm l$. Several orders of the satellite reflections arising from the $a_1/a_2$ domain patterns are observed around each SL reflection. The (103) reflection was chosen because the $a_1/a_2$ domain satellite peaks are not visible in the $Q_{[100]}–Q_{[001]}$ plane. The right panel of Fig.~\ref{fig:Xray}b is a line cut along the $\Delta Q_{[110]}$ direction (red dashed line in Fig.~\ref{fig:Xray}b left panel). Two orders of satellite peaks, labeled as $\pm m$, are associated with the $a_1/a_2$ domain modulation. 

For the $a_1/a_2$ and vortex mixed phase sample, the left panel of Fig. \ref{fig:Xray}c is the RSM in the $Q_{[001]}–Q_{[100]}$ plane around the DyScO$_3$ (DSO) (002) reflection. The RSM shows the DSO (002) reflection, along with multiple orders of (PbTiO$_3$)$_n$/(SrTiO$_3$)$_n$ SL reflections. All SL reflections split, suggesting a mixed phase domain pattern. The split peaks are labeled as FE $\pm l$ (from the $a_1/a_2$ domains), and V $\pm l$ (from the vortex domains). Several orders of the satellite reflections, arising from the vortex domain periodicity, are observed around each SL reflection. However, satellite peaks around the FE peaks are absent, since $a_1/a_2$ domain satellites are not visible in the $Q_{[100]}–Q_{[001]}$ plane. The right panel is a line cut around vortex $l=-1$ reflection along the $Q_{[100]}$ direction (red dashed line in Fig.~\ref{fig:Xray}c left panel). This line cut reveals two orders of satellite peaks, labeled as $\pm m$, associated with the vortex domain modulation. The peaks are absent in the line cut around the FE $l=-1$ reflection (blue dashed line in Fig.~\ref{fig:Xray}c left panel).

\section{Equilibrium and pump-induced SHG}\label{sec2}
\begin{figure*}[h!]
\centering
\includegraphics[width=1\textwidth]{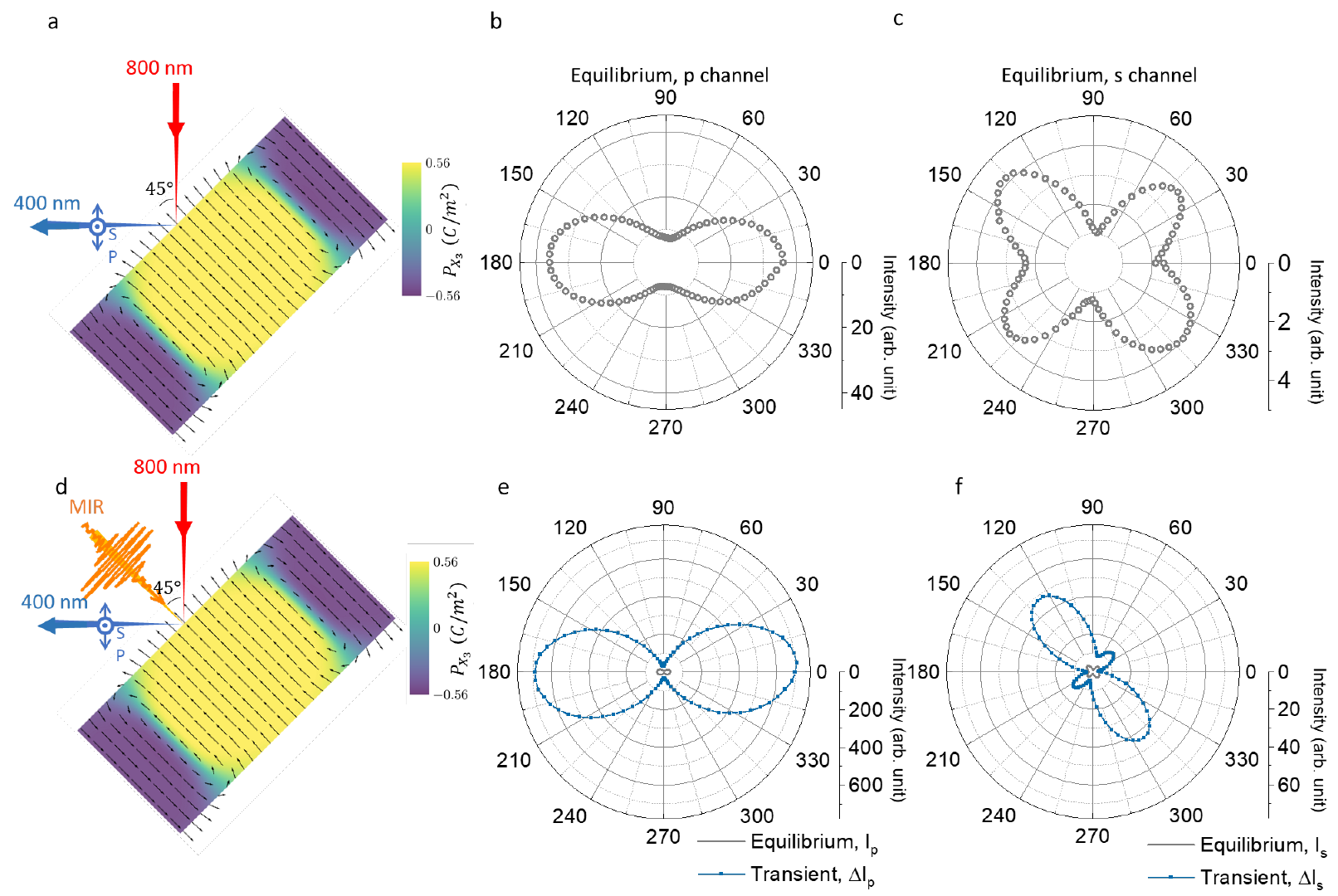}
\caption{\textbf{Dramatic light-induced change of SHG} \textbf{a,} The SHG polarimetry was measured by varying the polarization angle $\phi$ of the fundamental beam and with fixed detection polarization ($p$ or $s$). \textbf{b,} Equilibrium SHG polarimetry with $p$ detection. \textbf{c,} Equilibrium SHG polarimetry with $s$ detection. The $s$ channel SHG intensity is one order of magnitude smaller than that of the $p$ channel. \textbf{d,} The mid-infrared (MIR) field is polarized in the SHG incidence plane. \textbf{e,} Equilibrium (gray) and transient (blue) SHG polarimetry with $p$ detection. \textbf{f,} Equilibrium (gray) and transient (blue) SHG polarimetry with $s$ detection.} \label{fig:SHG}
\end{figure*}

In a simplified picture, the winding in-plane polarization components of the skyrmion largely cancel upon spatial averaging, while a finite out-of-plane polar component remains. This strongly suppresses the macroscopic in-plane second-harmonic generation (SHG) response, while the finite out-of-plane polar component can still contribute to the $p$-polarized equilibrium SHG signal at visible-light wavelengths. We conducted room-temperature SHG polarimetry measurements on a (PbTiO$_3$)$_{16}$/(SrTiO$_3$)$_{16}$ polar skyrmion sample in reflection geometry (Fig.~\ref{fig:SHG}a). The 45-degree incident angle allows us to optically access both the in- and out-of-plane information. The SHG intensity was measured along two orthogonal directions. The $s$ direction is perpendicular to the plane of incidence, and parallel to the superlattice’s surface. The $p$ direction is parallel to the plane of incidence. Overall, the equilibrium SHG intensity is small, with the $p$ channel signal about 10 times larger than that of the $s$ channel (Fig.~\ref{fig:SHG}b vs.~\ref{fig:SHG}c), consistent with the sensitivity of the $p$ channel to the finite out-of-plane polar response. In the transient state, the light-induced SHG intensity is about 20 times larger than the equilibrium value for both channels (Figs.~\ref{fig:SHG}e and ~\ref{fig:SHG}f).

\begin{figure*}
\centering
\includegraphics[width=1\textwidth]{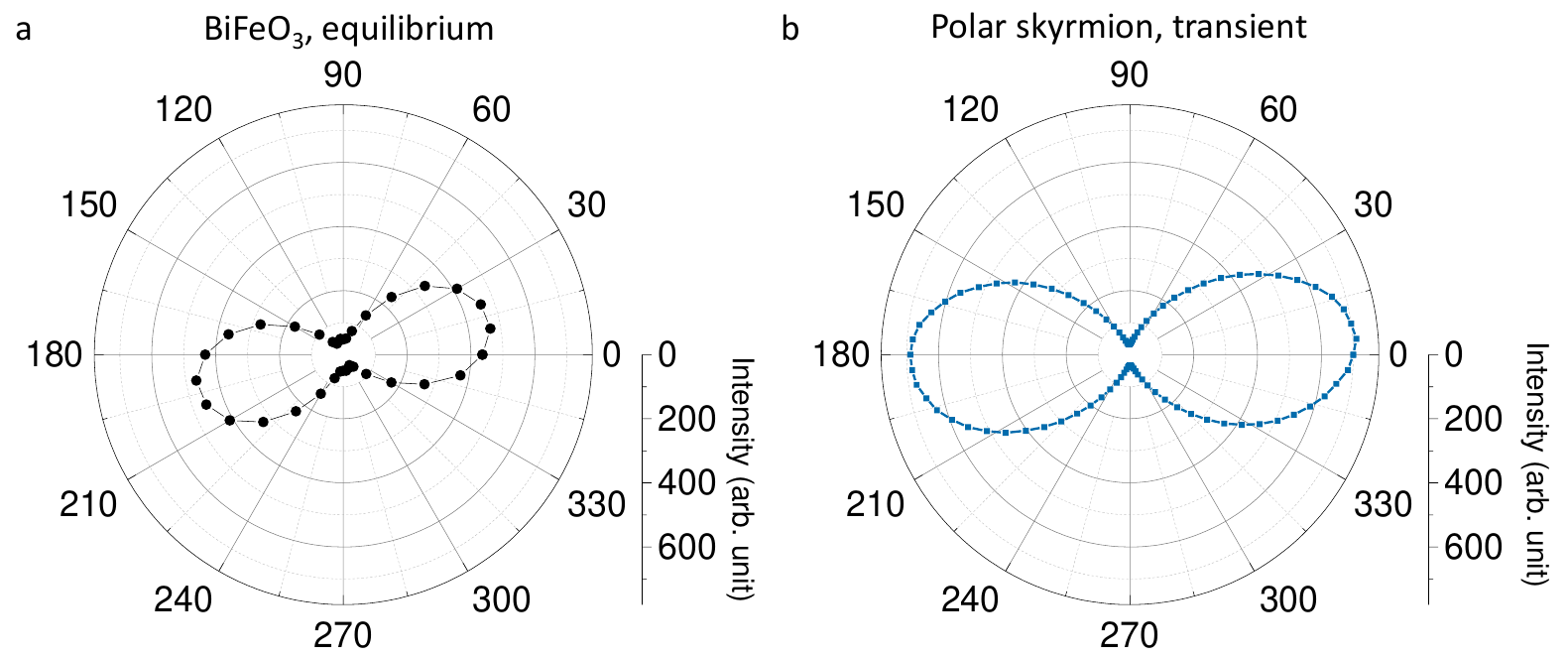}
\caption{\textbf{Giant transient SHG from the polar skyrmion.} \textbf{a,} Equilibrium SHG polarimetry of a single-domain BiFeO$_3$ with $p$ detection, measured under the same condition as the polar skyrmion, except for the absence of the mid-infrared excitation. \textbf{b,} The transient SHG polarimetry from the polar skyrmion.} \label{fig:BFO}
\end{figure*}

\begin{figure*}
\centering
\includegraphics[width=1\textwidth]{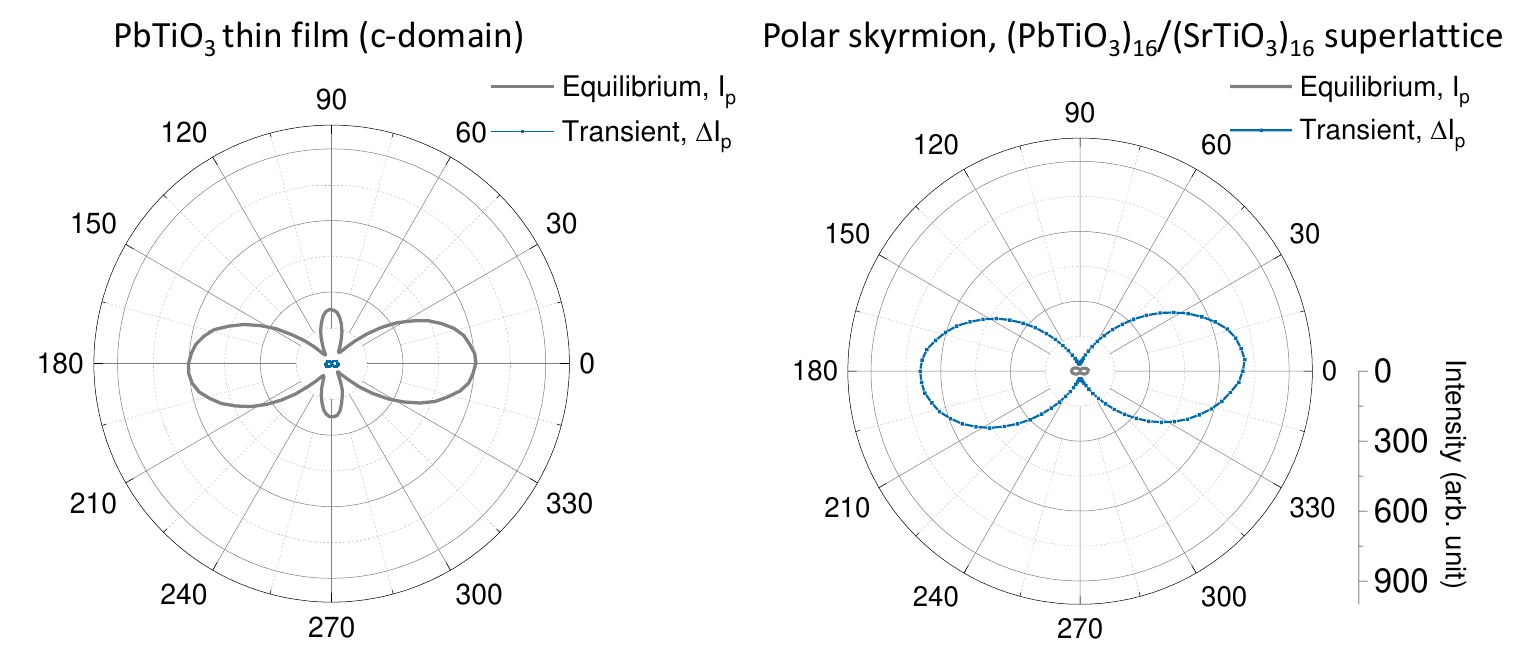}
\caption{\textbf{PbTiO$_3$ c-domain film versus skyrmion-hosting PbTiO$_3$/SrTiO$_3$ superlattices.} The non-superlattice film exhibits only a weak MIR-induced SHG response, in stark contrast to the giant transient SHG in the skyrmion-hosting superlattice.} \label{fig:PTOfilmvssuperlattice}
\end{figure*}

To demonstrate the giant MIR-induced transient SHG from the polar skyrmion sample, here we compare it with the equilibrium SHG polarimetry measured in the same setup (without MIR pumping) on a single-domain (001)$_{pc}$ BiFeO$_3$ (Fig.~\ref{fig:BFO}). We also compare the transient response from a 100-nm-thick PbTiO$_3$ film with out-of-plane ferroelectric polarization. With the same MIR excitation, the non-superlattice PbTiO$_3$ film exhibits a substantially weaker SHG intensity change than the skyrmion-hosting superlattice (Fig.~\ref{fig:PTOfilmvssuperlattice}).

\section{SHG polarimetry fit}\label{sec3}

In this section, we describe the symmetry model used to calculate the equilibrium and MIR-induced SHG response of the skyrmion phase. At room temperature, PbTiO$_3$ (PTO) is in the tetragonal phase with \textit{4mm} symmetry. We define a global sample coordinate system $(X,Y,Z)$, where $Z$ is normal to the film surface, $X$ lies in the SHG plane of incidence, and $Y$ is perpendicular to the plane of incidence. Because the skyrmion is a spatially varying polarization texture rather than a homogeneous domain, we distinguish this global sample frame from the local crystal axes $(X_1,X_2,X_3)$, where $X_3$ is chosen along the local polar axis and $X_1$ and $X_2$ span the plane perpendicular to $X_3$.

To describe the measured macroscopic SHG response, we approximate the skyrmion phase by an effective single-domain \textit{4mm} response with its polar axis along the sample normal $Z$. This is an effective symmetry description of the macroscopic optical response rather than a microscopic description of the full skyrmion texture. This approximation captures the leading macroscopic symmetry because the in-plane polarization components associated with the winding texture largely cancel upon spatial averaging, leaving an effective response predominantly characterized by the out-of-plane polar axis. We therefore choose $X_1\parallel X$, $X_2\parallel Y$, and $X_3\parallel Z$. In this effective crystal frame, the second-order nonlinear polarization can be written as

\begin{equation}
\mathbf{P}^{(2)} =
\begin{pmatrix}
P_1^{(2)} \\
P_2^{(2)} \\
P_3^{(2)}
\end{pmatrix}
=
\begin{pmatrix}
0 & 0 & 0 & 0 & d_{15} & 0 \\
0 & 0 & 0 & d_{15} & 0 & 0 \\
d_{31} & d_{31} & d_{33} & 0 & 0 & 0
\end{pmatrix}
\begin{pmatrix}
E_1^2 \\
E_2^2 \\
E_3^2 \\
2E_2E_3 \\
2E_1E_3 \\
2E_1E_2
\end{pmatrix}.
\label{eq:chi2}
\end{equation}

Here the indices $1$, $2$, and $3$ refer to the local crystal axes $X_1$, $X_2$, and $X_3$, respectively, $E_i$ are the components of the fundamental optical field in this frame, and $P_i^{(2)}$ are the corresponding components of the nonlinear polarization at the second-harmonic frequency. The coefficients $d_{15}$, $d_{31}$, and $d_{33}$ denote the independent nonzero components of the second-order susceptibility $\chi^{(2)}$ in contracted notation for \textit{4mm} symmetry.

\subsection{Fresnel coefficients}

To relate the nonlinear polarization inside the film to the detected SHG field, we account for reflection, transmission, and multiple reflections using Fresnel/local-field factors \cite{b4}. We represent the optical structure by an effective three-medium model in which media 1, 2, and 3 correspond to air, PTO, and the substrate, respectively. Since the $XZ$ plane is the plane of incidence, the $X$ and $Z$ components belong to the $p$-polarized field, whereas the $Y$ component is $s$ polarized.

At the fundamental optical frequency $\omega$, the Fresnel/local-field factors are

\begin{equation}
\begin{aligned}
L_{XX}(\omega) &=
\dfrac{t_{12}^p}{1+r_{12}^p r_{23}^p e^{2i\beta(\omega)}}
\left(1-r_{23}^p e^{2i\beta(\omega)}\right)
\dfrac{\cos\theta_2}{\cos\theta_1},
\\
L_{YY}(\omega) &=
\dfrac{t_{12}^s}{1+r_{12}^s r_{23}^s e^{2i\beta(\omega)}}
\left(1+r_{23}^s e^{2i\beta(\omega)}\right),
\\
L_{ZZ}(\omega) &=
\dfrac{t_{12}^p}{1+r_{12}^p r_{23}^p e^{2i\beta(\omega)}}
\left(1+r_{23}^p e^{2i\beta(\omega)}\right)
\dfrac{n_1(\omega)n_2(\omega)}
{n_{\mathrm{interface}}^2(\omega)}.
\end{aligned}
\end{equation}

The same expressions are evaluated at $2\omega$ for the outgoing SHG field. Here $n_i$ and $\theta_i$ are the refractive index and propagation angle in medium $i$, respectively, evaluated at the relevant optical frequency. The phase acquired by the wave while propagating through medium 2 is $\beta(\omega)=(2\pi/\lambda_\omega)n_2(\omega)d\cos\theta_2$, where $\lambda_\omega$ is the vacuum wavelength corresponding to the optical frequency $\omega$ and $d$ is the effective PTO thickness represented by medium 2 in this three-medium model. Following Ref.~\cite{b4}, with medium 2 taken as the nonlinear medium, $n_{\mathrm{interface}}=n_2$ for the incoming fundamental field and $n_{\mathrm{interface}}=n_1$ for the outgoing SHG field. The linear Fresnel reflection and transmission coefficients are

\begin{equation}
\begin{aligned}
r_{ij}^p &=
\dfrac{n_j\cos\theta_i-n_i\cos\theta_j}
{n_j\cos\theta_i+n_i\cos\theta_j},
&
r_{ij}^s &=
\dfrac{n_i\cos\theta_i-n_j\cos\theta_j}
{n_i\cos\theta_i+n_j\cos\theta_j},
\\
t_{ij}^p &=
\dfrac{2n_i\cos\theta_i}
{n_j\cos\theta_i+n_i\cos\theta_j},
&
t_{ij}^s &=
\dfrac{2n_i\cos\theta_i}
{n_i\cos\theta_i+n_j\cos\theta_j}.
\end{aligned}
\end{equation}

The input and output Fresnel factors are included by replacing each susceptibility component with its Fresnel-dressed effective value, $\chi_{ijk}^{(2)}\rightarrow L_{ii}^{(2\omega)}L_{jj}^{(\omega)}L_{kk}^{(\omega)}\chi_{ijk}^{(2)}$, where $\omega$ corresponds to the 800~nm probe.

For the $45^\circ$ incidence geometry, the fundamental probe field expressed in the sample frame has components $E_X=E_0\cos\phi/\sqrt{2}$, $E_Y=E_0\sin\phi$, and $E_Z=E_0\cos\phi/\sqrt{2}$, where $\phi=0^\circ$ corresponds to probe polarization in the SHG plane of incidence. Since the local crystal axes of the effective skyrmion model are aligned with the sample axes, substituting these field components into Eq.~\ref{eq:chi2} and including the input and output Fresnel factors gives the equilibrium SHG nonlinear polarization

\begin{equation}
\mathbf{P}^{(2)}_{\mathrm{samp}}
=
E_0^2
\begin{pmatrix}
L_{XX}^{(2\omega)}L_{XX}^{(\omega)}L_{ZZ}^{(\omega)}d_{15}\cos^2\phi
\\[4pt]
\dfrac{1}{\sqrt{2}}L_{YY}^{(2\omega)}L_{YY}^{(\omega)}L_{ZZ}^{(\omega)}d_{15}\sin(2\phi)
\\[4pt]
\dfrac{1}{2}L_{ZZ}^{(2\omega)}
\left[
\left(
\left(L_{ZZ}^{(\omega)}\right)^2 d_{33}
+
\left(L_{XX}^{(\omega)}\right)^2 d_{31}
\right)\cos^2\phi
+
2\left(L_{YY}^{(\omega)}\right)^2 d_{31}\sin^2\phi
\right]
\end{pmatrix}.
\label{eq:P2sk}
\end{equation}

The nonlinear polarization is then projected onto the $p$- and $s$-polarized detection channels. For the $45^\circ$ reflection geometry, we find

\begin{equation}
P_p^{(2)}=\dfrac{P_X^{(2)}-P_Z^{(2)}}{\sqrt{2}},
\qquad
P_s^{(2)}=P_Y^{(2)}.
\label{eq:P}
\end{equation}

The detected equilibrium SHG intensity is proportional to the squared magnitude of the corresponding projected nonlinear polarization, $I_\sigma\propto\lvert P_\sigma^{(2)}\rvert^2$, with $\sigma\in\{p,s\}$.

\subsection{$\chi^{(3)}$ process}\label{Sec:chi3}

In addition to the intrinsic second-order SHG response, the MIR electric field can generate a field-dependent nonlinear polarization through the third-order susceptibility. This contribution is linear in the MIR electric field. Since the MIR frequency $\Omega$ is much smaller than the optical probe frequency $\omega$, we treat the MIR field as slowly varying on the optical timescale and describe its effect as a time-dependent modulation of the nonlinear polarization near $2\omega$:

\begin{equation}
P_i^{(3)}(2\omega,t)
=
\chi_{ijkl}^{(3)}
E_j^{\mathrm{probe}}(\omega)
E_k^{\mathrm{probe}}(\omega)
E_l^{\mathrm{pump}}(t).
\label{eq:chi3_general}
\end{equation}

For convenience, for each pump-field direction $l$ we write the corresponding third-order susceptibility as a $3\times6$ matrix $\chi_l^{(3)}$ acting on the contracted quadratic probe-field vector. The third-order nonlinear polarization can then be written as

\begin{equation}
\mathbf{P}^{(3)}
=
\left(
E_1^{\mathrm{pump}}\chi_1^{(3)}
+
E_2^{\mathrm{pump}}\chi_2^{(3)}
+
E_3^{\mathrm{pump}}\chi_3^{(3)}
\right)
\begin{pmatrix}
(E_1^{\mathrm{probe}})^2 \\
(E_2^{\mathrm{probe}})^2 \\
(E_3^{\mathrm{probe}})^2 \\
2E_2^{\mathrm{probe}}E_3^{\mathrm{probe}} \\
2E_1^{\mathrm{probe}}E_3^{\mathrm{probe}} \\
2E_1^{\mathrm{probe}}E_2^{\mathrm{probe}}
\end{pmatrix}.
\label{eq:chi3_contracted}
\end{equation}

For \textit{4mm} symmetry, the corresponding matrices for the three pump-field directions are

\begin{equation}
\begin{aligned}
\chi^{(3)}_1 &=
\begin{pmatrix}
d_{11}^{(1)} & d_{12}^{(1)} & d_{13}^{(1)} & 0 & 0 & 0 \\
0 & 0 & 0 & 0 & 0 & d_{26}^{(1)} \\
0 & 0 & 0 & 0 & d_{35}^{(1)} & 0
\end{pmatrix},
\\[4pt]
\chi^{(3)}_2 &=
\begin{pmatrix}
0 & 0 & 0 & 0 & 0 & d_{26}^{(1)} \\
d_{12}^{(1)} & d_{11}^{(1)} & d_{13}^{(1)} & 0 & 0 & 0 \\
0 & 0 & 0 & d_{35}^{(1)} & 0 & 0
\end{pmatrix},
\\[4pt]
\chi^{(3)}_3 &=
\begin{pmatrix}
0 & 0 & 0 & 0 & d_{15}^{(3)} & 0 \\
0 & 0 & 0 & d_{15}^{(3)} & 0 & 0 \\
d_{31}^{(3)} & d_{31}^{(3)} & d_{33}^{(3)} & 0 & 0 & 0
\end{pmatrix}.
\end{aligned}
\label{eq:chi3matrices}
\end{equation}

The subscripts 1, 2, and 3 on $\chi_l^{(3)}$ denote the direction of the MIR electric field along the local crystal axes $X_1$, $X_2$, and $X_3$, respectively. In the effective skyrmion model, these axes coincide with the sample axes $X$, $Y$, and $Z$.

For clarity, Fresnel factors are omitted in the following higher-order symmetry analysis. At fixed probe wavelength and geometry, they renormalize the effective coefficients but do not alter the angular functional forms derived below.

Because the MIR pump is normally incident, its electric field lies in the sample plane. We first consider a MIR field polarized along $X=X_1$, i.e., parallel to the SHG plane of incidence. Using the probe-field components defined above gives

\begin{equation}
\mathbf{P}^{(3)}(\phi)
=
E^{\mathrm{pump}}
\left(E_0^{\mathrm{probe}}\right)^2
\begin{pmatrix}
\dfrac{1}{2}\left(d_{11}^{(1)}+d_{13}^{(1)}\right)\cos^2\phi+d_{12}^{(1)}\sin^2\phi
\\[4pt]
\dfrac{1}{\sqrt{2}}d_{26}^{(1)}\sin(2\phi)
\\[4pt]
d_{35}^{(1)}\cos^2\phi
\end{pmatrix}.
\label{eq:P3X}
\end{equation}

After projection onto the detector channels, the angular dependences can therefore be written as

\begin{equation}
\begin{aligned}
P_p^{(3)}(\phi) &\propto A\cos^2\phi+B\sin^2\phi,
\\
P_s^{(3)}(\phi) &\propto \sin(2\phi),
\end{aligned}
\label{eq:P3Xps}
\end{equation}

where $A$ and $B$ are effective coefficients containing the relevant tensor and optical factors.

For a MIR field polarized along $Y=X_2$, perpendicular to the SHG plane of incidence, the fourfold symmetry gives

\begin{equation}
\begin{aligned}
P_p^{(3)}(\phi) &\propto \sin(2\phi),
\\
P_s^{(3)}(\phi) &\propto C\cos^2\phi+D\sin^2\phi,
\end{aligned}
\label{eq:P3Yps}
\end{equation}

where $C$ and $D$ are the corresponding effective coefficients.

\subsection{$\chi^{(4)}$ process}\label{Sec:chi4}

We also consider a higher-order contribution that is quadratic in the MIR electric field and is described by the fourth-order nonlinear susceptibility,

\begin{equation}
P_i^{(4)}(2\omega,t)
=
\chi_{ijk,lm}^{(4)}
E_j^{\mathrm{probe}}(\omega)
E_k^{\mathrm{probe}}(\omega)
E_l^{\mathrm{pump}}(t)
E_m^{\mathrm{pump}}(t).
\label{eq:chi4_general}
\end{equation}

For a MIR field polarized along $X=X_1$, only the $l=m=1$ pump-field component contributes, so the expression reduces to $P_i^{(4)}=(E^{\mathrm{pump}})^2\chi_{ijk,11}^{(4)}E_j^{\mathrm{probe}}E_k^{\mathrm{probe}}$. For \textit{4mm} symmetry, the relevant contracted tensor can be written as

\begin{equation}
\left(\chi_{ijk,11}^{(4)}\right)
=
\begin{pmatrix}
0 & 0 & 0 & 0 & d_{15}^{(11)} & 0 \\
0 & 0 & 0 & d_{24}^{(11)} & 0 & 0 \\
d_{31}^{(11)} & d_{32}^{(11)} & d_{33}^{(11)} & 0 & 0 & 0
\end{pmatrix}.
\label{eq:chi4matrix}
\end{equation}

Using the same probe geometry gives

\begin{equation}
\mathbf{P}^{(4)}(\phi)
=
\left(E^{\mathrm{pump}}\right)^2
\left(E_0^{\mathrm{probe}}\right)^2
\begin{pmatrix}
d_{15}^{(11)}\cos^2\phi
\\[4pt]
\dfrac{1}{\sqrt{2}}d_{24}^{(11)}\sin(2\phi)
\\[4pt]
\dfrac{1}{2}\left(d_{31}^{(11)}+d_{33}^{(11)}\right)\cos^2\phi+d_{32}^{(11)}\sin^2\phi
\end{pmatrix}.
\label{eq:P4X}
\end{equation}

The corresponding detector-channel dependences are therefore

\begin{equation}
\begin{aligned}
P_p^{(4)}(\phi) &\propto A'\cos^2\phi+B'\sin^2\phi,
\\
P_s^{(4)}(\phi) &\propto \sin(2\phi),
\end{aligned}
\label{eq:P4Xps}
\end{equation}

where $A'$ and $B'$ are effective coefficients.

For a MIR field polarized along $Y=X_2$, the corresponding response follows from the $90^\circ$ rotational symmetry of the \textit{4mm} crystal. The tensor coefficients are transformed accordingly, while the detector-channel angular dependences retain the same functional form,

\begin{equation}
\begin{aligned}
P_p^{(4)}(\phi) &\propto C'\cos^2\phi+D'\sin^2\phi,
\\
P_s^{(4)}(\phi) &\propto \sin(2\phi),
\end{aligned}
\label{eq:P4Yps}
\end{equation}

with $C'$ and $D'$ denoting the corresponding effective coefficients.

\section{Pump polarization dependence}\label{sec4}
\begin{figure*}[h]
\centering
\includegraphics[width=1\textwidth]{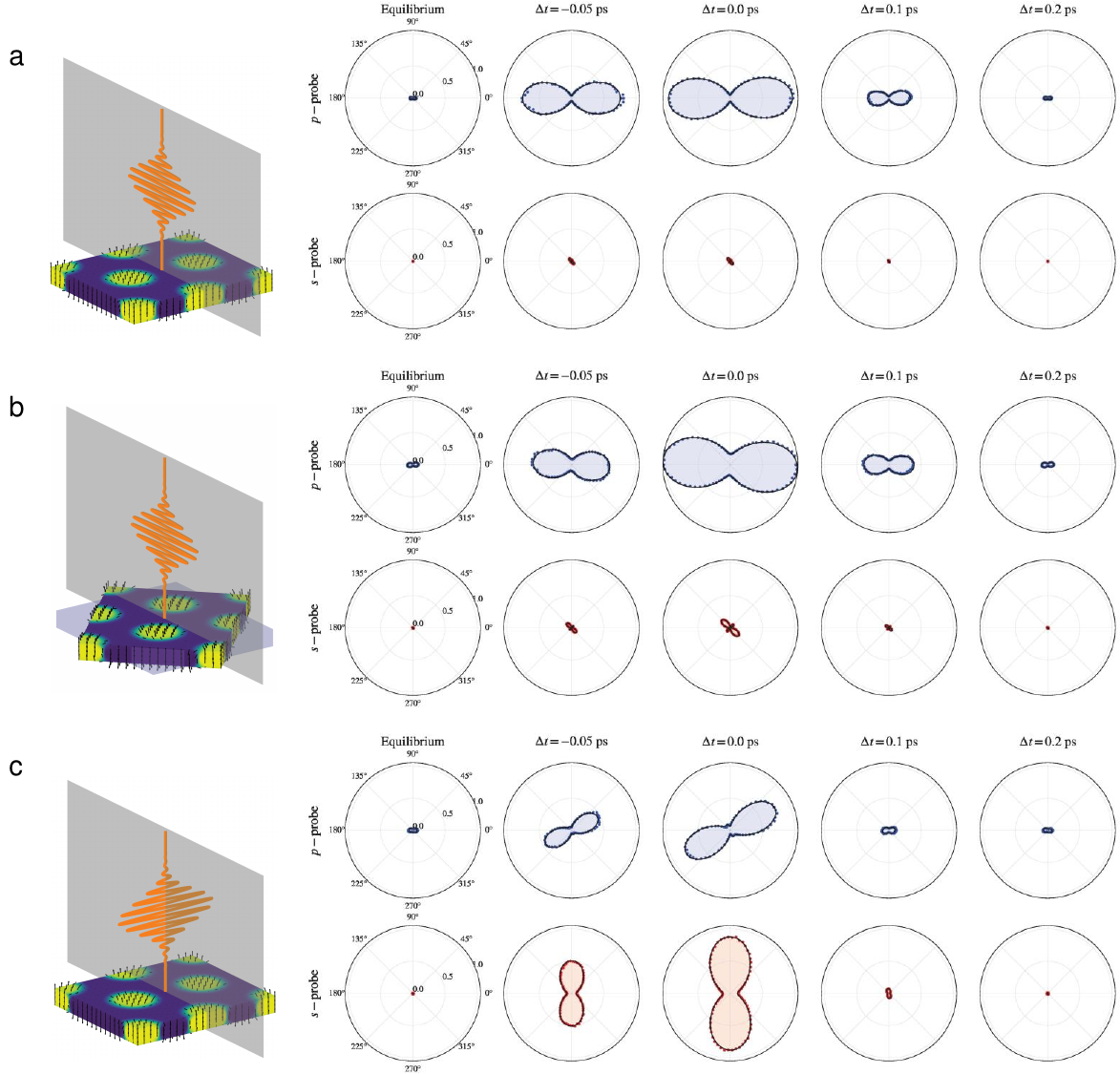}
\caption{\textbf{Pump polarization and sample orientation dependence.} Transient SHG polarimetry probed in two orthogonal channels at $t=0$ ps with three different geometries: \textbf{a,} the initial geometry defined in main text Fig.1a and Fig.~\ref{fig:SHG}a. \textbf{b,} sample rotated by 45$^\circ$ along the surface normal, and \textbf{c,} MIR polarization rotated 90$^\circ$. The SHG intensity in all cases are normalized by the maximum transient intensity $\Delta I_p$ in case \textbf{a}.} \label{fig:polarizationDep}
\end{figure*}

\section{Polar texture dependence}\label{sec5}
\begin{figure*}[h]
\centering
\includegraphics[width=0.9\textwidth]{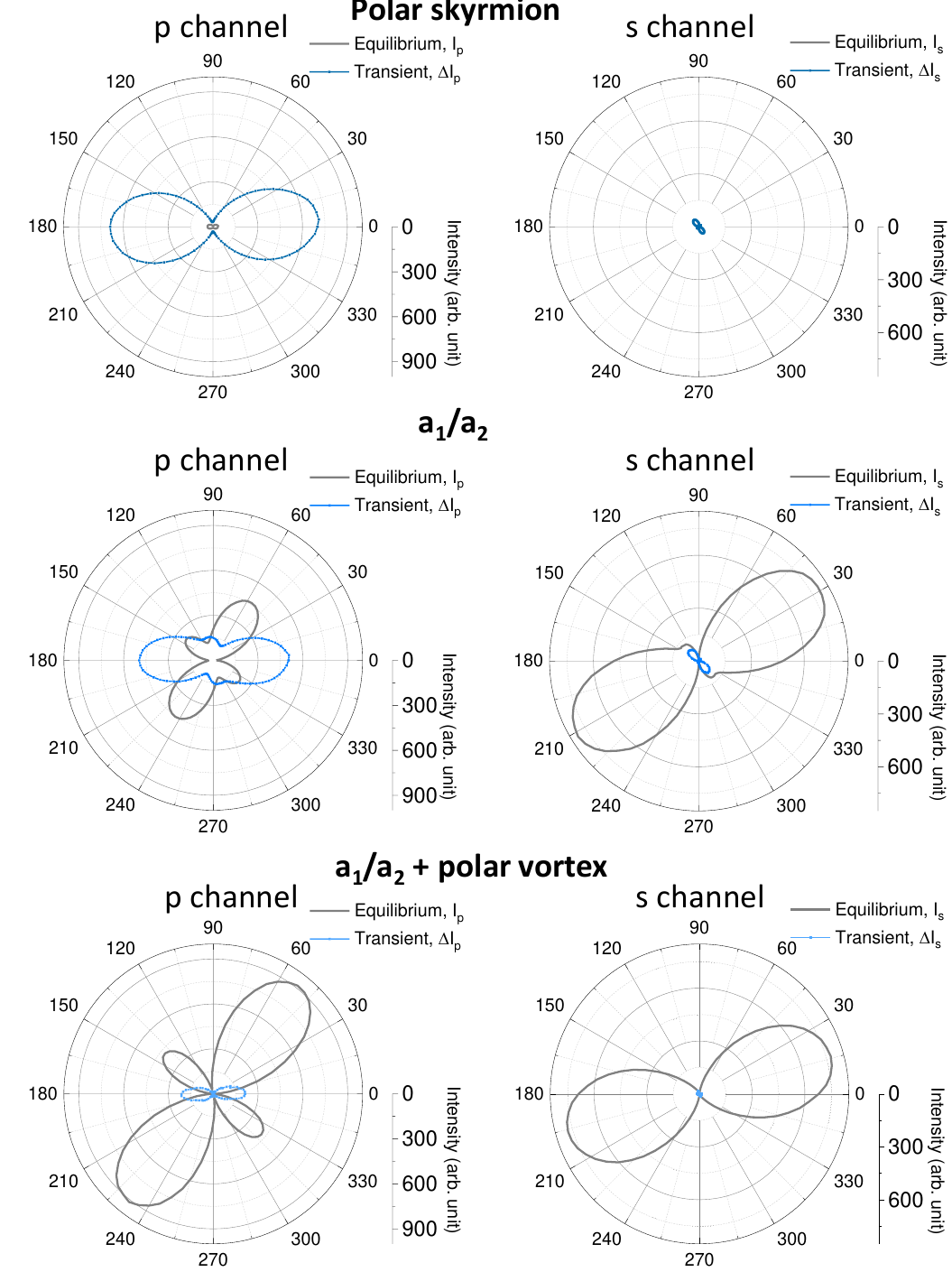}
\caption{\textbf{Polar texture dependence.} Equilibrium versus transient SHG for the polar skyrmion, the a$_1$/a$_2$ sample, and the mixed phase sample. Here the MIR polarization is horizontal (along 0$^\circ$). For all samples, the transient responses at $t=0$ps are shown.} \label{fig:SampleDepS}
\end{figure*}

All polar phases show a strong light-induced SHG, as plotted in Fig.~\ref{fig:SampleDepS}. The transient SHG intensity from the polar skyrmion phase is about 3 times stronger than those from the a$_1$/a$_2$ and the mixed phase samples.  

\section{Absence of THz emission}\label{sec6}

\begin{figure*}[h]
\centering
\includegraphics[width=1\textwidth]{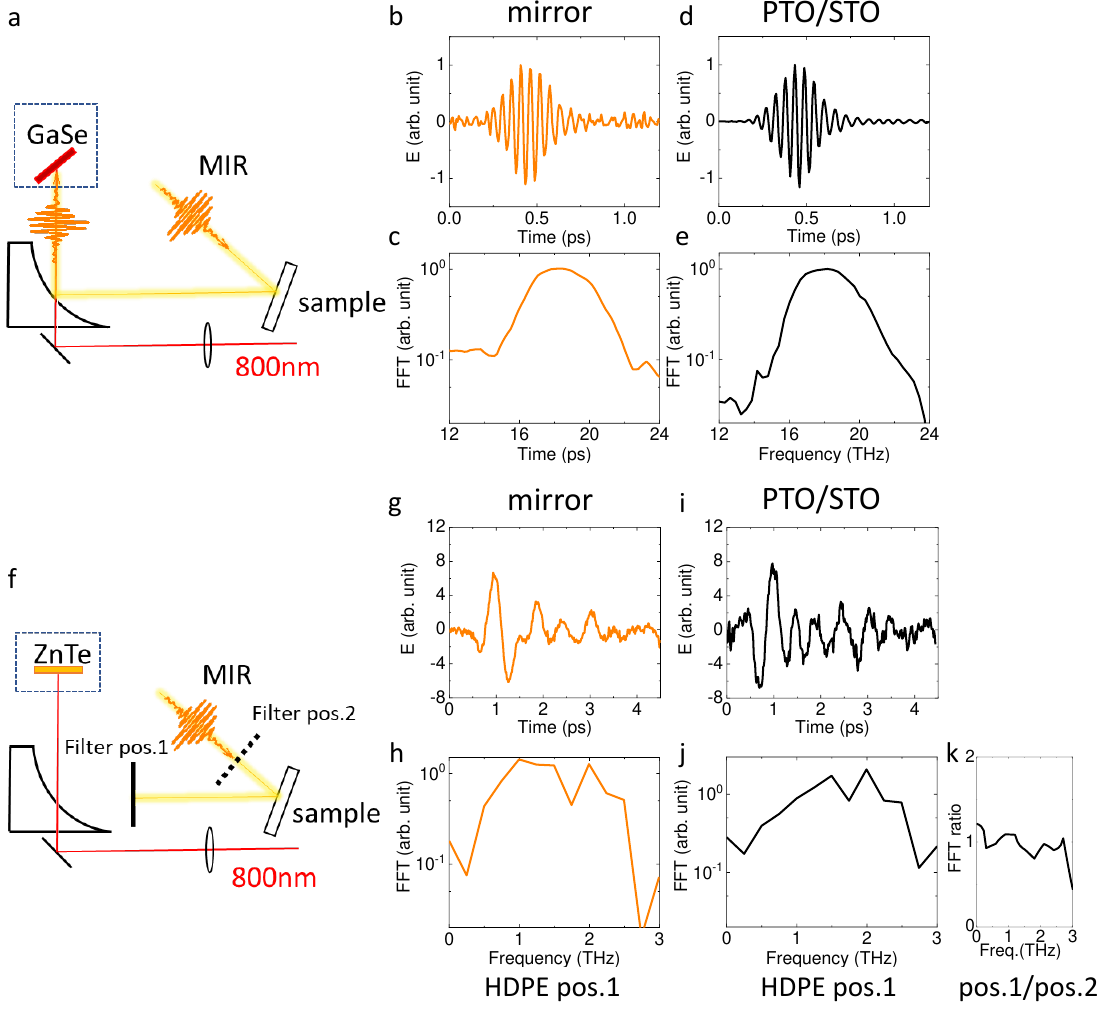}
\caption{\textbf{Characterization of the excitation pulse and potential optical rectification.} \textbf{a,} Electro-optical (EO) sampling in a 0.2-mm thick GaSe to characterize the MIR excitation. Either a gold mirror or the polar skyrmion sample was placed at the sample position. The MIR incident angle is 22.5$^\circ$. EO sampling of the gold-mirror-reflected MIR in the time domain (\textbf{b}) and in the frequency domain (\textbf{c}). EO sampling result of the MIR after interacting with the polar skyrmion: time-domain result (\textbf{d}) and normalized FFT amplitude (\textbf{e}). \textbf{f,} THz emission setup. A MIR filter was placed either after the sample (pos.1) or before the sample (pos.2). \textbf{g,} A weak THz signal was detected when using the gold mirror. \textbf{h,} Normalized FFT amplitude. \textbf{i,} A weak THz signal nearly identical to the gold-mirror case was detected from the polar skyrmion, indicating the absence of additional THz emission from the polar skyrmion. \textbf{j,} Normalized FFT amplitude. \textbf{k,} A near unity FFT ratio of the THz signal is seen, with the filter behind the sample versus in front of the sample. Therefore, the THz signal is from the MIR generation, but not from optical rectification of the MIR. All measurements were done in air without purging.  The strong oscillations in the time domain after the main THz pulse and the features in the frequency domain are due to water absorption.} \label{fig:EOsampling}
\end{figure*}

We have conducted electro-optical sampling measurement to characterize the wavelength of the MIR pump and to investigate potential pump-induced optical rectification. The results are shown in Fig.\ref{fig:EOsampling}. A weak THz signal is seen from the polar skyrmion (Figs.~\ref{fig:EOsampling}i, ~\ref{fig:EOsampling}j). However, This THz signal is identical to the one measured with a gold mirror (Figs.~\ref{fig:EOsampling}g, ~\ref{fig:EOsampling}h) instead of the polar skyrmion. In addition, when placing a MIR filter in front or behind the sample, the THz signal shows no difference (Fig.~\ref{fig:EOsampling}k). These observations indicate that the THz signal is a residual from the mid-infrared generation. There is no far-field THz emission generated by the sample.

Based on the simplified 4$mm$ model (see the $\chi^{(2)}$ tensor as shown in Eq.\ref{eq:chi2}), we obtain the polarization induced by optical rectification of the pump field as: 
\begin{equation}\label{eq:MIRX1}
    \textbf{P(0)}\vert_{\textbf{E}_{MIR}\parallel X_1} = \begin{pmatrix} 0 & 0 & 0 & 0 & d_{15} & 0 \\ 0 & 0 & 0 & d_{15} & 0 & 0 \\ d_{31} & d_{31} & d_{33} & 0 & 0 & 0 \end{pmatrix} \begin{pmatrix} E_{MIR}^2 \\ 0 \\ 0 \\ 0 \\ 0 \\ 0 \end{pmatrix}
= \begin{pmatrix}0 \\ 0 \\ d_{31}E^2_{MIR} \end{pmatrix}
\end{equation}
\begin{equation}\label{eq:MIRX2}
    \textbf{P(0)}\vert_{\textbf{E}_{MIR}\parallel X_2} =\begin{pmatrix} 0 & 0 & 0 & 0 & d_{15} & 0 \\ 0 & 0 & 0 & d_{15} & 0 & 0 \\ d_{31} & d_{31} & d_{33} & 0 & 0 & 0 \end{pmatrix} \begin{pmatrix} 0 \\ E_{MIR}^2 \\ 0 \\ 0 \\ 0 \\ 0 \end{pmatrix}
= \begin{pmatrix}0 \\ 0 \\ d_{31}E^2_{MIR} \end{pmatrix}
\end{equation}
Both cases give the same result on optical rectification induced polarization.

\section{Heterodyned and homodyned SHG components}\label{sec8}
For processes which involve the third-order nonlinearity, the pump electric field $E(\omega_1)$ brings in an additional contribution to the second-harmonic polarization $P(2\omega\pm\omega_1)\propto \chi^{(3)} E(\omega_1)E(\omega)E(\omega)$. If the material has a nonzero $\chi^{(2)}$, the SHG intensity is given by
\begin{equation} \label{eq:chi2chi3}
I\propto\Bigl\{  \left[\chi^{(2)}+\chi^{(3)}E(\omega_1)\right]E(\omega)E(\omega) \Bigr\}  ^2
\end{equation}
This equation has three terms: the equilibrium SHG intensity, a heterodyned term oscillates at the pump frequency $\omega_1$ (observed in non-centrosymmetric materials\cite{VonHoegen2018, Grishunin2017}), and a homodyned term containing $\left[\chi^{(3)}E(\omega_1)\right]^2$ which includes a displacive response at $2\omega$ and a $2\omega_1$ oscillatory term (observed in centrosymmetric systems with zero $\chi^{(2)}$\cite{Grishunin2017}). All these terms are summarized in Table~\ref{tabchi2chi3}.

\begin{table}[h]
\caption{Heterodyned and homodyned SHG components involving $\chi^{(2)}$ and $\chi^{(3)}$.}\label{tabchi2chi3}%
\begin{tabular}{@{}llll@{}}
\toprule
Term & Description  & Frequency & Pump field dependence\\
\midrule
$\left[\chi^{(2)}E(\omega)E(\omega)\right]^2\equiv E^2_0$    & Equilibrium    & $2\omega$  & N/A  \\
2$\chi^{(2)}\chi^{(3)}E(\omega_1)\left[E(\omega)E(\omega)\right]^2\equiv 2E_0E(\chi^{(3)})$    & Heterodyne   & $2\omega\pm\omega_1$ \footnotemark[1]  & $E(\omega_1)$  \\
$\left[\chi^{(3)}E(\omega_1)E(\omega)E(\omega)\right]^2\equiv E^2(\chi^{(3)})$    & Homodyne  & $2\omega$,$2\omega\pm2\omega_1$\footnotemark[2]   & $E^2(\omega_1)$ \\
\botrule
\end{tabular}
\footnotetext[1]{This corresponds to a modulation of the transient SHG intensity at the pump frequency $\omega_1$.}
\footnotetext[2]{This is the $2\omega_1$ oscillation. In our measurement, this term is filtered out by the bandpass filter.}
\end{table}

Following the notation in Table~\ref{tabchi2chi3}, we now discuss the processes involving the fourth-order optical susceptibility. In this case, the second-harmonic generation intensity is given by
\begin{equation}
\label{eq:chi2chi3chi4}
I\propto\Bigl\{  \left[\chi^{(2)}+\chi^{(3)}E(\omega_1)+\chi^{(4)}E(\omega_1)E(\omega_1)\right]E(\omega)E(\omega)\Bigl\} ^2
\end{equation}
All terms contributing to the SHG intensity are listed in Table~\ref{tabchi2chi3chi4}. Oscillatory components at frequencies higher than the pump frequency ($\omega_1$) are removed by the bandpass filter in front of the photomultiplier. This table reveals that the quasi-static component of the signal depends on $E^2(\omega_1)$ and $E^4(\omega_1)$, and that the amplitude of the oscillation at $\omega_1$ exhibits dependence on $E(\omega_1)$ and $E^3(\omega_1)$.

Alternatively, the third term in Eq.~\ref{eq:chi2chi3chi4}, $\chi^{(4)}E(\omega_1)E(\omega_1)E(\omega)E(\omega)$, can be replaced by a cascaded $\chi^{(2)}$ and $\chi^{(3)}$ process, $\chi^{(3)}E(\omega)E(\omega)P(0)$ with $P(0)=\chi^{(2)}E(\omega_1)E(-\omega_1)$. Therefore, the $\chi^{(4)}$ term in Table~\ref{tabchi2chi3chi4} corresponds to either an intrinsic or an effective $\chi^{(4)}$ process.

For an effective $\chi^{(4)}$ generated via cascaded optical rectification and a subsequent $\chi^{(3)}$ process, the rectification-induced polarization field remains perpendicular to the sample plane upon a 90$^\circ$ pump polarization rotation (see Eqs.~\ref{eq:MIRX1},\ref{eq:MIRX2}). Thus, the $E^2(\chi^{(4)})$ term in Table~\ref{tabchi2chi3chi4} remains independent of the pump polarization. In contrast, the $E^2(\chi^{(3)})$ term varies with the pump polarization, which explains the experimental observation (Fig.~\ref{fig:polarizationDep}).

\begin{table}[h]
\caption{Heterodyned and homodyned SHG components involving $\chi^{(2)}$, $\chi^{(3)}$ and $\chi^{(4)}$. The $2\omega$ and $2\omega\pm\omega_1$ terms are measured, while higher frequency sidebands are removed by the bandpass filter.}\label{tabchi2chi3chi4}%
\begin{tabular}{@{}llll@{}}
\toprule
Term & Description  & Frequency & Pump field dependence\\
\midrule
$E^2_0$    & Equilibrium    & $2\omega$  & N/A  \\
2$E_0E(\chi^{(3)})$    & Heterodyne   & $2\omega\pm\omega_1$   & $E(\omega_1)$  \\
2$E_0E(\chi^{(4)})$    & Heterodyne   & $2\omega$,$2\omega\pm2\omega_1$   & $E^2(\omega_1)$  \\
2$E(\chi^{(3)})E(\chi^{(4)})$    & Heterodyne   & $2\omega\pm\omega_1$, $2\omega\pm3\omega_1$  & $E^3(\omega_1)$  \\
$E^2(\chi^{(3)})$    & Homodyne  & $2\omega$,$2\omega\pm2\omega_1$ & $E^2(\omega_1)$ \\
$E^2(\chi^{(4)})$    & Homodyne  & $2\omega$,$2\omega\pm2\omega_1$,$2\omega\pm4\omega_1$ & $E^4(\omega_1)$ \\
\botrule
\end{tabular}
\end{table}

\begin{figure*}[h]
\centering
\includegraphics[width=0.7\textwidth]{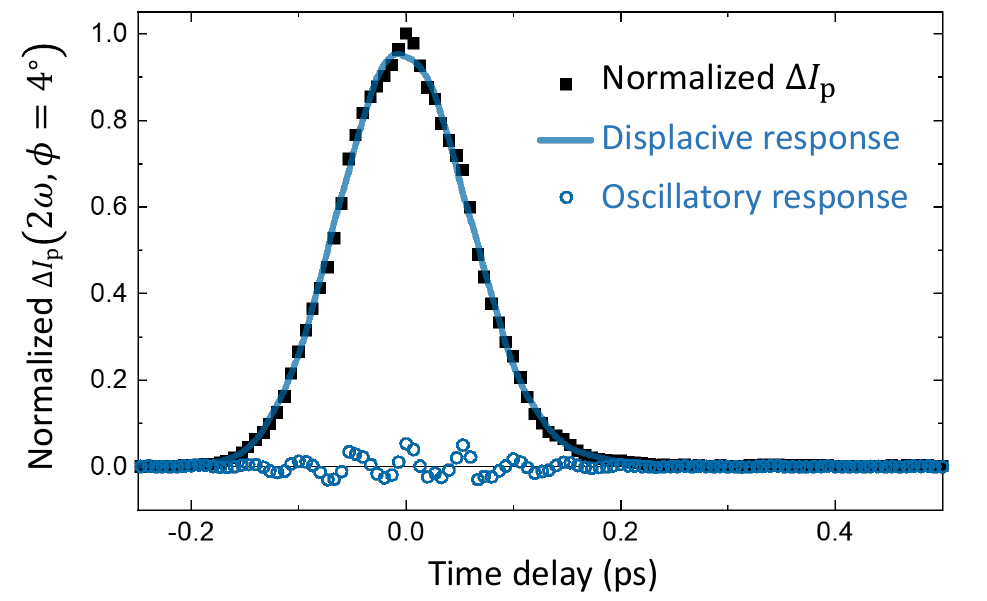}
\caption{\textbf{Extraction of displacive and oscillatory responses from experimental data.} With the MIR pumping, the transient SHG intensity of the polar skyrmion sample shows a giant enhancement with oscillations at the pump frequency. The non-oscillatory contribution is modeled using a Gaussian function and identified as the displacive response. The oscillatory response is obtained by subtracting this fitted displacive component from the raw data.} \label{fig:FitOverlap}
\end{figure*}

In the experiment, we measured the change in second-harmonic generation (SHG) intensity as a function of the pump–probe time delay. A representative pump–probe trace of the polar skyrmion sample is shown in Fig.~\ref{fig:FitOverlap}. The transient SHG intensity exhibits a pronounced enhancement accompanied by oscillations at the pump frequency. The displacive response is defined as the non-oscillatory component of the transient signal, which is modeled using a Gaussian function
\begin{equation}
\label{eq:gaussian}
y(x)=A\exp\left[-\left(\frac{x-x_0}{\sqrt2\sigma}\right)^2\right] 
\end{equation}
where $A$ is the peak amplitude, $x_0$ is the peak timing, and $\sigma$ is the standard deviation of the Gaussian distribution. The time traces were acquired over multiple scans. Each scan was independently fitted with a Gaussian function, and the standard deviation of the extracted peak displacive response across all scans was taken as the uncertainty for a given pump fluence. The oscillatory component was obtained by subtracting the averaged displacive response from the averaged raw time-trace at the same pump fluence. We then evaluated the Fourier transform amplitude of the oscillatory component. The oscillation amplitude was extracted from the area of the corresponding frequency-domain peak using a Gaussian function (Eq.~\ref{eq:gaussian}) in the frequency domain, where $x_0$ is the peak center frequency. The peak area is $A\sigma\sqrt{2\pi}$. The uncertainty in $A$ is estimated via standard error propagation, including contributions from spectral background noise, Fourier-transform noise, and the amplitude fitting uncertainty derived from the corresponding confidence interval. The uncertainty in $\sigma$ is estimated from the 90$\%$ confidence interval of the fit.

\begin{figure*}[h]
\centering
\includegraphics[width=1\textwidth]{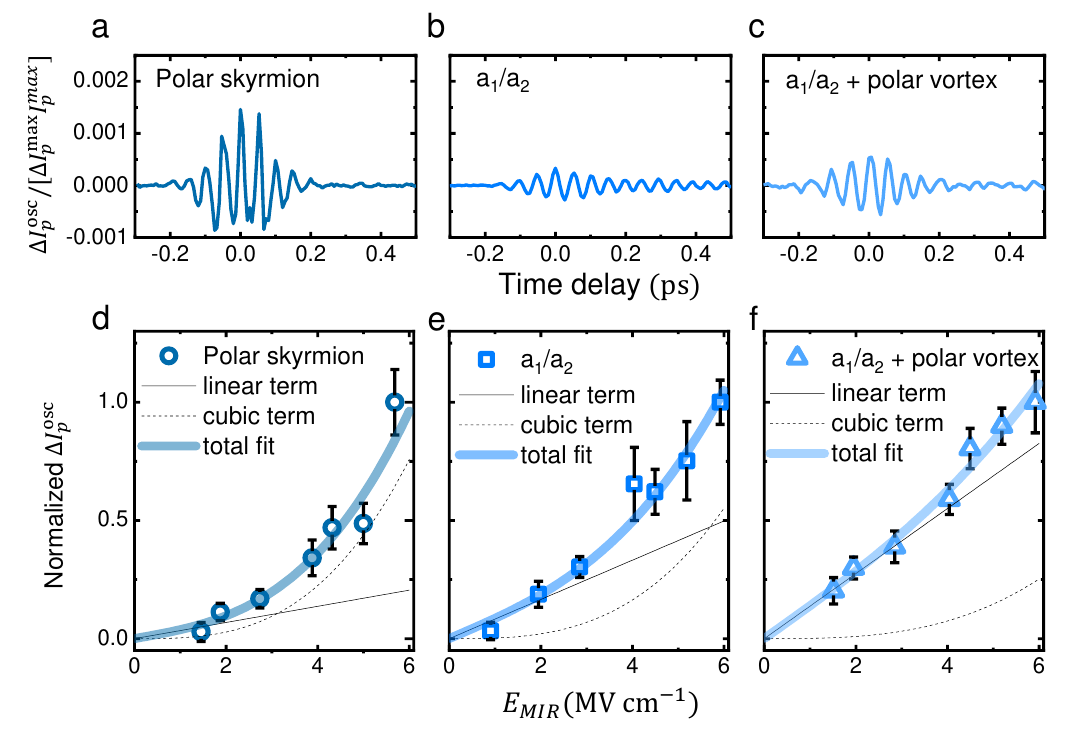}
\caption{\textbf{Evidence of the $\chi^{(3)}$ and $\chi^{(4)}$ contributions to the oscillatory response.} The transient SHG intensity oscillates at the pump frequency, with a combined linear and cubic pump-field dependence. This figure shows the same data as Fig.~4 in the main text, with the individual contributions of the linear (solid line) and cubic (dashed line) terms to the oscillatory response highlighted. Fitting coefficients for $\Delta I^{\mathrm{osc}} = a E_{\mathrm{MIR}} + b E_{\mathrm{MIR}}^3$ are shown in Table~\ref{tabcoefficients}.} \label{fig:fitcomparison}
\end{figure*}

Table~\ref{tabchi2chi3chi4} shows that the $\chi^{(3)}$ contributes to a transient SHG intensity oscillation at the pump frequency that is linearly proportional to the pump field. In addition, $\chi^{(3)}$ and $\chi^{(4)}$ contribute to a cubic pump-field dependence of the oscillation at the pump frequency. Therefore, analyzing the pump-field dependence of the SHG intensity oscillation provides access to higher-order optical susceptibilities. Figure~\ref{fig:fitcomparison} presents the same data as Fig.~4 in the main text and highlights the individual contributions of the linear and cubic terms to the oscillatory response. The $\chi^{(4)}$ contribution becomes prominent at high pump fluences. Table~\ref{tabcoefficients} shows the fitting parameters.       

\begin{table}[h]
\caption{Fitting coefficients for $\Delta I^{\mathrm{osc}} = a E_{\mathrm{MIR}} + b E_{\mathrm{MIR}}^3$.}\label{tabcoefficients}%
\begin{tabular}{@{}lll@{}}
\toprule
polar texture & a  & b \\
\midrule
polar skyrmion    & 0.034205    & 0.003506 \\
$a_1/a_2$    & 0.082982   & 0.002556 \\
$a_1/a_2$ + polar vortex    & 0.137605  & 0.001169 \\
\botrule
\end{tabular}
\end{table}
In addition, we have tested whether the pump-field dependence for three polar textures is better described by a single global mixed order scaling model ($aE_{MIR}+bE_{MIR}^3$) over the entire measured field range, or a two-regime (piecewise) description is preferred. We used the small-sample correct Akaike information criterion (AICc) as the primary model-selection metric. In addition to model selection, we computed diagnostics from the global mixed fits to summarize the relative importance of the low- and high-order terms and the stability of inferred crossover scales. The conclusions are:\\
1. The oscillatory responses for the polar skyrmion phase and the $a_1/a_2$ phase are both well described by a single global mixed odd-order model ($aE_{MIR}+bE_{MIR}^3$) over the measured field range. In both cases, piecewise alternatives are not preferred after AICc penalization, and this conclusion remains stable across the tested weighting scan.\\
2. The oscillatory response of the mixed $a_1/a_2$ and polar vortex phase is best described by a simpler single-$E$ global model. The cubic contribution remains weak over the measured range, the mixed-model crossover is pushed beyond the measured field range, and piecewise alternatives are not competitive.

\section{Pump wavelength dependence}\label{sec8}
\begin{figure*}[h]
\centering
\includegraphics[width=0.6\textwidth]{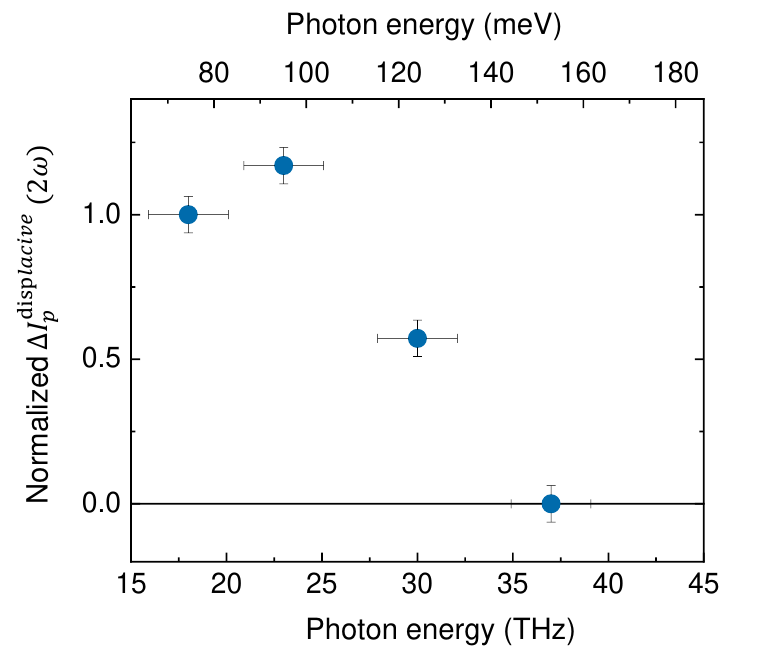}
\caption{\textbf{Pump wavelength dependence.}  The normalized displacive response in $\Delta I_p(2\omega)$ as a function of the excitation photon energy. Vertical error bars represent the standard deviation. Horizontal bars are the bandwidths of the mid-infrared pump pulses.} \label{fig:FreqDep}
\end{figure*}

The peak response of the transient SHG intensity versus pump photon energy for the polar skyrmion is shown in Fig.~\ref{fig:FreqDep}. According to Ref.\cite{Freire1988}, the highest A$_1$ phonon in PbTiO$_3$ is located at 646 cm$^{-1}$ (or 19.4 THz, TO frequency), for which our MIR pump could potentially resonantly excite. However, the transient SHG does not peak at this frequency, but is close to the LO phonon frequency\cite{Freire1988}.

\section{Phase-field simulations}\label{sec9}

The phase-field formulation used to model the PTO/STO superlattice defines the free energy density as follows:

\begin{equation}
\begin{aligned}
    f =& \alpha_i P_i^2 + \alpha_{ij}P_i^2P_j^2 + \alpha_{ijk}P_i^2P_j^2P_k^2 +\dfrac{1}{2}c_{ijkl}\varepsilon_{ij}\varepsilon_{kl} - q_{ijkl}\varepsilon_{ij}P_kP_l\\
    &+\dfrac{1}{2}G_{ijkl}P_{i,j}P_{k,l} -\dfrac{1}{2}K_{ij}E_iE_j -E_iP_i.
\end{aligned}
\end{equation}

In this expression, $\alpha_i$, $\alpha_{ij}$, and $\alpha_{ijk}$ represent the Landau coefficients accounting for the thermodynamic expansion of the free energy; $c_{ijkl}$ are the elastic stiffness parameters; $q_{ijkl}=c_{ijmn}Q_{mnkl}$, where $Q_{ijkl}$ is the electrostrictive tensor, capture the electrostrictive coupling between strain and polarization; $G_{ijkl}$ are associated with the polarization gradient energy; and $K_{ij}=\kappa^b\delta_{ij}$ denote the background dielectric permittivity tensor components. The variables $P_i$, $\varepsilon_{ij}$, $E_i$ ,and $u_i$ correspond to the components of polarization, mechanical strain, electric field, and displacement, respectively. 

The elastic strain is given by $\varepsilon_{ij}^{\mathrm{elas}}=\varepsilon_{ij}-\varepsilon_{ij}^0$, where $\varepsilon_{ij}$ is the total strain and $\varepsilon_{ij}^0=Q_{ijkl}P_kP_l$ is the eigen strain. The total strain is composed of two parts: $\varepsilon_{ij}=\varepsilon_{ij}^{\mathrm{ext}}+\eta_{ij}$, where $\varepsilon_{ij}^{\mathrm{ext}}$ is the homogeneous strain (the external mechanical strain), and $\eta_{ij}=\dfrac{1}{2}\left(u_{i,j}+u_{j,i}\right)$ is the inhomogeneous strain. The total electric field $E_i$ includes the effective THz field from our pump and the internal fields. Summation is implied over repeated indices, following the Einstein convention across spatial dimensions $x_i$ ($i = 1, 2, 3$), and partial derivatives with respect to spatial coordinates are denoted by subscript “$,i$”.

The temporal evolution of the polarization field is governed by the polarization dynamic equation:\cite{Akamatsu2018,Yang2020}

\begin{equation}
\mu \dfrac{\partial^2 P_i}{\partial t^2} + \gamma \dfrac{\partial P_i}{\partial t} = -\dfrac{\delta F}{\delta P_i}
\end{equation}

Here, $\mu$ is the effective inertial coefficient of polarization, $\gamma$ is the damping constant, and $F = \int f dV$ represents the total free energy over the simulated volume\cite{QianLi2021}.
In addition, the system satisfies both the mechanical equilibrium condition:
\begin{equation}
    \sigma_{ij,j} = \dfrac{\partial}{\partial x_j}\left( \dfrac{\partial f}{\partial \varepsilon_{ij}} \right) = 0
\end{equation}
and the electrostatic Maxwell’s equation:
\begin{equation}
    D_{i,i} = -\dfrac{\partial}{\partial x_i}\left( \dfrac{\partial f}{\partial E_i}\right) = 0.
\end{equation}
These conditions hold under the assumptions of a body-force-free and charge-free ferroelectric medium, where $\sigma_{ij}$ and $D_i$ are the stress and electric displacement components, respectively. To numerically solve the resulting coupled partial differential equations, a semi-implicit Fourier-spectral method \cite{Chen1998} is employed, offering high computational efficiency and accuracy in handling spatial derivatives.

The material parameters used for PTO and STO in the simulations are taken from references \cite{Haun1987,Sheng2010,QianLi2021,Tan2021}, from which we take an effective polarization mass density $\mu$ of $6.27 \times 10^{-18}$ $\mathrm{kg} \cdot \mathrm{m}^3 / \mathrm{C}^2$ and a damping coefficient $\gamma$ of $2\times 10^{-7}$ $\mathrm{\Omega}\cdot\mathrm{m}$. This effective polarization mass density reproduces the frequencies of the PbTiO$_3$ E(TO) and A$_1$(TO) modes. To initiate the simulation, a small random fluctuation in polarization—less than $0.01 P_0$, where $P_0 = 0.76$ $\mathrm{C}/\mathrm{m}^2$ is the spontaneous polarization of PTO at room temperature—is introduced as the initial condition. The simulations were done for room temperature (293 K).

\begin{table}[h]

\begin{center}
\caption{Parameters used for simulating three phases.}\label{tabPFSimulation}
\begin{tabular}{c c c} 

 \hline
 Coefficient & PTO & STO  \\ [0.5ex] 
 \hline

 \multicolumn{3}{c}{Landau coefficient (Temperature $T$ in K)} \\
  \hline

$\alpha_1$ ($10^7$ $\mathrm{C}^{-2}\mathrm{m}^2\mathrm{N}$) & $0.038(T - 752) $ & $4.05[\coth(54/T) - 1.056]$ \\
$\alpha_{11}$ ($10^8$ $\mathrm{C}^{-4}\mathrm{m}^6\mathrm{N}$) & $-0.73$ & $1.7$ \\
$\alpha_{12}$ ($10^9$ $\mathrm{C}^{-4}\mathrm{m}^6\mathrm{N}$) & $0.75$ & $4.45$ \\
$\alpha_{111}$ ($10^8$ $\mathrm{C}^{-4}\mathrm{m}^{10}\mathrm{N}$)  & $2.6$ & \\
$\alpha_{112}$ ($10^8$ $\mathrm{C}^{-4}\mathrm{m}^{10}\mathrm{N}$)  & $6.1$ & \\
$\alpha_{123}$ ($10^9$ $\mathrm{C}^{-4}\mathrm{m}^{10}\mathrm{N}$)  & $-3.7$ & \\
\hline
 \multicolumn{3}{c}{Gradient energy coefficient} \\

  \hline
$G_{11}$ ($10^{-10}$ $\mathrm{C}^{-2}\mathrm{m}^4\mathrm{N}$) & $1.038$ & $1.038$ \\
$G_{12}$ ($10^{-10}$ $\mathrm{C}^{-2}\mathrm{m}^4\mathrm{N}$) & $0$ & $0$ \\
$G_{44}$ ($10^{-10}$ $\mathrm{C}^{-2}\mathrm{m}^4\mathrm{N}$) & $0.519$ & $0.519$ \\
  \hline

 \multicolumn{3}{c}{Elastic stiffness} \\

   \hline
$c_{11}$ ($10^9$ Pa) & $230$ & $330$ \\
$c_{12}$ ($10^9$ Pa) & $100$ & $100$ \\
$c_{44}$ ($10^9$ Pa) & $70$ & $125$ \\
  \hline
  
 \multicolumn{3}{c}{Electrostrictive coupling coefficient} \\

    \hline
$q_{11}$ ($10^{10}$ $\mathrm{C}^{-2}\mathrm{m}^2\mathrm{N}$) & $1.142$ & $1.26$ \\
$q_{12}$ ($10^{9}$ $\mathrm{C}^{-2}\mathrm{m}^2\mathrm{N}$) & $0.49$ & $-13$ \\
 $q_{44}$ ($10^{9}$ $\mathrm{C}^{-2}\mathrm{m}^2\mathrm{N}$) & $3.718$ & $11.96$ \\

 $Q_{11}$ ($\mathrm{C}^{-2}\mathrm{m}^4$) & $0.089$ & $0.0457$ \\
$Q_{12}$ ($\mathrm{C}^{-2}\mathrm{m}^4$) & $-0.026$ & $-0.0135$ \\
 $Q_{44}$ ($\mathrm{C}^{-2}\mathrm{m}^4$) & $0.0675$ & $0.00957$ \\
     \hline

 \multicolumn{3}{c}{Background dielectric constant ($\epsilon_0 = 8.85\times 10^{-12}$ $ \mathrm{F}\mathrm{m}^{-1}$)} \\
\hline 
$\kappa^b$ ($\epsilon_0$) & $10$ & $10$ \\
\hline 

\end{tabular}

\end{center}
\end{table}

\begin{figure*}[h]
\centering
\includegraphics[width=1\textwidth]{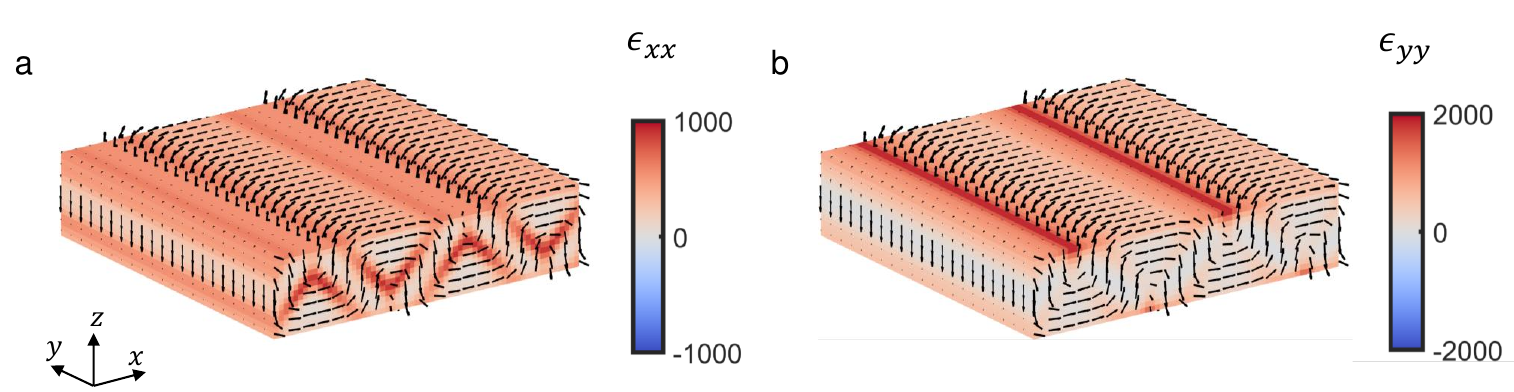}
\caption{\textbf{Phase-field simulations of the local permittivity distribution for the polar vortex phase.} Ferroelectric polarization distribution (arrows) and the local permittivity (color coded) of the PTO layer for the pure vortex phase. \textbf{a}, The in-plane permittivity, $\epsilon_{xx}$. \textbf{b}), The in-plane permittivity, $\epsilon_{yy}$.} \label{fig:vortex}
\end{figure*}

To simulated three polar phases, we used the same coefficients listed in Table~\ref{tabPFSimulation} for the free energy density. The only parameter varied was the applied biaxial strain. For the a$_1$/a$_2$ and skyrmion phases, we assumed that STO is strain-free, while PTO experiences a biaxial strain of $\epsilon_{xx} = \epsilon_{yy} = 0\%$ for the $a_1$/$a_2$ phase and $\epsilon_{xx} = \epsilon_{yy} = -1.34\%$ for the skyrmion phase. In the case of the vortex phase, the strain is anisotropic. For STO, the strain components are $\epsilon_{xx} = 1.2\%$ and $\epsilon_{yy} = 1.05\%$, while for PTO they are $\epsilon_{xx} = -0.12\%$ and $\epsilon_{yy} = -0.275\%$. We simulated the pure vortex phase for simplicity. We evaluated the in-plane permittivity for two cases, with an in-plane electric field applied perpendicular to (Fig.~\ref{fig:vortex}a) or parallel to the vortex core (Fig.~\ref{fig:vortex}b). For both cases, the local permittivity is predominantly positive, in contrast to the large alternating positive and negative local permittivity in the polar skyrmion case. As the domain boundaries are where large local permittivity arises, we anticipate the $a_1/a_2$ and vortex mixed phase to have a larger effective permittivity than either the pure vortex and the pure $a_1/a_2$ phase.

%% file: sn-bibliography.bib
@article{Franken1961,
  author		= "P. A. Franken and A. E. Hill and C. W. Peters and G. Weinreich",
  title			= "Generation of Optical Harmonics",
  journal		= "Phys. Rev. Lett.",
  volume		= "7",
  pages			= "118--119",
  year			= "1961",
  doi			= "10.1103/PhysRevLett.7.118"
}

@article{Wang2024,
  author		= "Sixu Wang and Wei Li and Chenguang Deng and Zijian Hong and Han-Bin Gao and Xiaolong Li and Yueliang Gu and Qiang Zheng and Yongjun Wu and Paul G. Evans and Jing-Feng Li and Ce-Wen Nan and Qian Li",
  title			= "Giant electric field-induced second harmonic generation in polar skyrmions",
  journal		= "Nature Commun.",
  volume		= "15",
  pages			= "1374",
  year			= "2024",
  doi = {10.1038/s41467-024-45755-5}
}

@article{Yadav2019,
  author		= "Ajay K. Yadav and Kayla X. Nguyen and Zijian Hong and Pablo García-Fernández and Pablo Aguado-Puente5 and
Christopher T. Nelson and Sujit Das and Bhagwati Prasad and Daewoong Kwon and Suraj Cheema and Asif I. Khan and Chenming Hu and Jorge Íñiguez and Javier Junquera and Long-Qing Chen and David A. Muller and Ramamoorthy Ramesh and Sayeef Salahuddin1",
  title			= "Spatially resolved steady-state negative capacitance",
  journal		= "Nature",
  volume		= "565",
  pages			= "468--471",
  year			= "2019",  
  doi = "10.1038/s41586-018-0855-y"
}

@article{Das2021,
  author		= "S. Das and Z. Hong and V. A. Stoica and M. A. P. Gonçalves and Y. T. Shao and E. Parsonnet and E. J. Marksz and S. Saremi and M. R. McCarter and A. Reynoso and C. J. Long and A. M. Hagerstrom and D. Meyers and V. Ravi and B. Prasad and H. Zhou and Z. Zhang and H. Wen and F. Gómez-Ortiz and P. García-Fernández and J. Bokor and J. Íñiguez and J. W. Freeland and N. D. Orloff and J. Junquera and L. Q. Chen and S. Salahuddin and D. A. Muller and L. W. Martin and R. Ramesh",
  title			= "Local negative permittivity and topological phase transition in polar skyrmions",
  journal		= "Nature Mater.",
  volume		= "20",
  pages			= "194--201",
  year			= "2021",
  doi ="10.1038/s41563-020-00818-y"
}

@article{Stoica2019,
  author		= "V. A. Stoica and N. Laanait and C. Dai and Z. Hong and Y. Yuan and Z. Zhang and S. Lei and M. R. McCarter and A. Yadav and A. R. Damodaran and S. Das and G. A. Stone and J. Karapetrova and D. A. Walko and X. Zhang and L. W. Martin and R. Ramesh and L.-Q. Chen and H. Wen and V. Gopalan and J. W. Freeland",
  title			= "Optical creation of a supercrystal with three-dimensional nanoscale periodicity",
  journal		= "Nature Mater.",
  volume		= "18",
  pages			= "377-–383",
  year			= "2019 ", 
  doi = "10.1038/s41563-019-0311-x"
}

@article{QianLi2021,
  author		= "Qian Li and Vladimir A. Stoica and Marek Paściak and Yi Zhu and Yakun Yuan and Tiannan Yang and Margaret R. McCarter and Sujit Das and Ajay K. Yadav and Suji Park and Cheng Dai and Hyeon Jun Lee and Youngjun Ahn and Samuel D. Marks and Shukai Yu and Christelle Kadlec and Takahiro Sato and Matthias C. Hoffmann and Matthieu Chollet and Michael E. Kozina and Silke Nelson and Diling Zhu and Donald A. Walko and Aaron M. Lindenberg and Paul G. Evans and Long-Qing Chen and Ramamoorthy Ramesh and Lane W. Martin and Venkatraman Gopalan and John W. Freeland and Jirka Hlinka and Haidan Wen",
  title			= "Subterahertz collective dynamics of polar vortices",
  journal		= "Nature",
  volume		= "592",
  pages			= "376–-380",
  year			= "2021 ",
  doi = "10.1038/s41586-021-03342-4"
}

@article{Damodaran2017,
  author		= "A. R. Damodaran and J. D. Clarkson and Z. Hong and H. Liu and A. K. Yadav and C. T. Nelson and S.-L. Hsu and M. R. McCarter and K.-D. Park and V. Kravtsov and A. Farhan and Y. Dong and Z. Cai and H. Zhou and P. Aguado-Puente and P. García-Fernández and J. Íñiguez and J. Junquera and A. Scholl and M. B. Raschke and L.-Q. Chen and D. D. Fong and R. Ramesh and L. W. Martin",
  title			= "Phase coexistence and electric-field control of toroidal order in oxide superlattices",
  journal		= "Nature Mater.",
  volume		= "16",
  pages			= "1003–-1009",
  year			= "2017",
  doi = "10.1038/nmat4951"
}

@article{Yadav2016,
  author		= "A. K. Yadav and C. T. Nelson and S. L. Hsu and Z. Hong and J. D. Clarkson and C. M. Schlepütz and A. R. Damodaran and P. Shafer and E. Arenholz and L. R. Dedon and D. Chen and A. Vishwanath and A. M. Minor and L. Q. Chen and J. F. Scott and L. W. Martin and R. Ramesh",
title ="Observation of polar vortices in oxide superlattices",
  journal		= "Nature",
  volume		= "530",
  pages			= "198--201",
  year			= "2016",
  doi = "10.1038/nature16463"
}

@article{Das2019,
  author		= "S. Das and Y. L. Tang and Z. Hong and M. A. P. Gonçalves and M. R. McCarter and C. Klewe and K. X. Nguyen and F. Gómez-Ortiz and P. Shafer and E. Arenholz and V. A. Stoica and S.-L. Hsu and B. Wang and C. Ophus and J. F. Liu and C. T. Nelson and S. Saremiand B. Prasad and A. B. Mei and D. G. Schlom and J. Íñiguez and P. García-Fernández and D. A. Muller and L. Q. Chen and J. Junquera and L. W. Martin and R. Ramesh",
  title ="Observation of room-temperature polar skyrmions",
  journal		= "Nature",
  volume		= "568",
  pages			= "368--372",
  year			= "2019",
  doi ="10.1038/s41586-019-1092-8"
}

@book{YRShenBook,
  author		= "Y. R. Shen",
  title			= "The {P}rinciples of {N}onlinear {O}ptics",
  address		= "New Jersey",
  publisher		= "Wiley-Interscience",
  year			= "2002"
}

@book{BoydBook,
    author = {Robert W. Boyd},
    title = {{Nonlinear Optics}},
    address = "New York",
    year = {2008},
    publisher = {Academic Press}
}

@article{Sionnest1987,
  author		= "P. Guyot-Sionnest and J. H. Hunt and Y. R. Shen",
  title ="Sum-Frequency Vibrational Spectroscopy of a Langmuir Film: Study of Molecular Orientation of a Two-Dimensional System",
  journal		= "Phys. Rev. Lett.",
  volume		= "59",
  pages			= "1597--1600",
  year			= "1987",
  doi			= "10.1103/PhysRevLett.59.1597"
}

@article{Manzoni2016,
  author		= "C. Manzoni and G. Cerullo",
  title ="Design criteria for ultrafast optical parametric amplifiers",
  journal		= "J. Opt.",
  volume		= "18",
  pages			= "103501",
  year			= "2016",
  doi			= "10.1088/2040-8978/18/10/103501"
}

@article{Yue2024,
  author		= "Li Yue and Chang Liu and Shanshan Han and Hao Hong and Yijun Wang and Qiaomei Liu and Jiajie Qi and Yuan Li and Dong Wu and Kaihui Liu and Enge Wang and Tao Dong and Nanlin Wang",
  title ="Giant nonlinear optical wave mixing in a van der Waals correlated insulator",
  journal		= "Sci. Adv.",
  volume		= "10",
  pages			= "eadn6216",
  year			= "2024",
  doi			= "10.1126/sciadv.adn6216"
}

@article{New1967,
  author		= "G. H. C. New and J. F. Ward",
  title ="Optical Third-Harmonic Generation in Gases",
  journal		= "Phys. Rev. Lett.",
  volume		= "19",
  pages			= "556-–559",
  year			= "1967",
  doi			= "10.1103/PhysRevLett.19.556"
}

@article{Hendry2010,
  author		= "E. Hendry and P. J. Hale and J. Moger and A. K. Savchenko and S. A. Mikhailov",
  title ="Coherent Nonlinear Optical Response of Graphene",
  journal		= "Phys. Rev. Lett.",
  volume		= "105",
  pages			= "097401",
  year			= "2010",
  doi			= "10.1103/PhysRevLett.105.097401"
}

@article{Kauranen2012,
  author		= "Martti Kauranen and Anatoly V. Zayats",
  title ="Nonlinear plasmonics",
  journal		= "Nat. Photonics",
  volume		= "6",
  pages			= "737",
  year			= "2012",
  doi			= "10.1038/NPHOTON.2012.244"
}

@article{Pettine2024,
  author		= "Jacob Pettine and Prashant Padmanabhan and Teng Shi and Lauren Gingras and Luke McClintock and Chun-Chieh Chang and Kevin W. C. Kwock and Long Yuan and Yue Huang and John Nogan and Jon K. Baldwin and Peter Adel and Ronald Holzwarth and Abul K. Azad and Filip Ronning and Antoinette J. Taylor and Rohit P. Prasankumar and Shi-Zeng Lin and Hou-Tong Chen ",
  title =  "Light-driven nanoscale vectorial currents",
  journal		= "Nature",
  volume		= "626",
  pages			= "984-–989",
  year			= "2024",
  doi			= "10.1038/s41586-024-07037-4"
}

@article{Denev2011,
  author		= "Sava A. Denev and Tom T. A. Lummen and Eftihia Barnes and Amit Kumar and Venkatraman Gopalan",
  title =  "Probing Ferroelectrics Using Optical Second Harmonic Generation",
  journal		= "J. Am. Ceram. Soc.",
  volume		= "94",
  pages			= "2699–-2727",
  year			= "2011",
  doi			= "10.1111/j.1551-2916.2011.04740.x"
}

@article{Fiebig2005,
  author		= "Manfred Fiebig and Victor V. Pavlov and Roman V. Pisarev",
  title =  "Second-harmonic generation as a tool for studying electronic and magnetic structures of crystals: review",
  journal		= "J. Opt. Soc. Am. B",
  volume		= "22",
  pages			= "96–-117",
  year			= "2005",
  doi			= "10.1364/JOSAB.22.000096"
}

@article{Hafez2018,
  author		= "Hassan A. Hafez and Sergey Kovalev and Jan-Christoph Deinert and Zoltán Mics and Bertram Green and Nilesh Awari and Min Chen and Semyon Germanskiy and Ulf Lehnert and Jochen Teichert and Zhe Wang and Klaas-Jan Tielrooij and Zhaoyang Liu and Zongping Chen and Akimitsu Narita and Klaus Müllen and Mischa Bonn and Michael Gensch and Dmitry Turchinovich",
  title ="Extremely efficient terahertz high-harmonic generation in graphene by hot Dirac fermions",
  journal		= "Nature",
  volume		= "561",
  pages			= "507-–511 ",
  year			= "2018",
  doi			= "10.1038/s41586-018-0508-1"
}

@article{Junquera2023,
  author		= "Javier Junquera and Yousra Nahas and Sergei Prokhorenko and Laurent Bellaiche and Jorge Íñiguez and Darrell G. Schlom and Long-Qing Chen and Sayeef Salahuddin and David A. Muller and Lane W. Martin and R. Ramesh",
  title ="Topological phases in polar oxide nanostructures",
  journal		= "Rev. Mod. Phys.",
  volume		= "95",
  pages			= "025001",
  year			= "2023",
  doi			= "10.1103/RevModPhys.95.025001"
}

@article{Zubko2016,
  author		= "Pavlo Zubko and Jacek C. Wojdeł and Marios Hadjimichael and Stéphanie Fernandez-Pena and Anaïs Sené and Igor Luk’yanchuk and Jean-Marc Triscone and Jorge Íñiguez",
  title ="Negative capacitance in multidomain ferroelectric superlattices",
  journal		= "Nature",
  volume		= "534",
  pages			= "524--528",
  year			= "2016",
  doi			= "10.1038/nature17659"
}

@article{Sun2025,
  author		= "Haoying Sun and Pengcheng Chen and Wei Mao and Changqing Guo and Yueying Li and Jierong Wang and Wenjie Sun and Duo Xu and Bo Hao and Tingjun Zhang and Jianan Ma and Jiangfeng Yang and Zhequan Cao and Shengjun Yan and Yuze Guan and Zonghan Wen and Zhangwen Mao and Ningchong Zheng and Zhengbin Gu and Houbing Huang and Peng Wang and Yong Zhang and Di Wu and Yuefeng Nie",
  title ="Ferroelectric topologies in {BaTiO$_3$} nanomembranes for light field manipulation",
  journal		= "Nat. Nanotechnol",
  volume		= "534",
  pages			= "524--528",
  year			= "2025",
  doi			= "10.1038/s41565-025-01919-y"
}

@article{Disa2021,
  author		= "Ankit S. Disa and Tobia F. Nova and Andrea Cavalleri",
  title ="Engineering crystal structures with light",
  journal		= "Nat. Phys.",
  volume		= "17",
  pages			= "1087--1092",
  year			= "2021",
  doi			= "10.1038/s41567-021-01366-1"
}

@article{Subedi2015,
  author		= "Alaska Subedi",
  title ="Proposal for ultrafast switching of ferroelectrics using midinfrared pulses",
  journal		= "Phys. Rev. B",
  volume		= "92",
  pages			= "214303",
  year			= "2015",
  doi			= "10.1103/PhysRevB.92.214303"
}

@article{Mankowsky2017,
  author		= "R. Mankowsky and A. von Hoegen and M. Först and A. Cavalleri",
  title ="Ultrafast Reversal of the Ferroelectric Polarization",
  journal		= "Phys. Rev. Lett.",
  volume		= "118",
  pages			= "197601",
  year			= "2017",
  doi			= "10.1103/PhysRevLett.118.197601"
}

@article{Zhang2024,
  author		= "Xinshu Zhang and Tyler Carbin and Adrian B. Culver and Kai Du and Kefeng Wang and Sang-Wook Cheong and Rahul Roy and Anshul Kogar",
  title ="Light-induced electronic polarization in antiferromagnetic {Cr$_2$O$_3$}",
  journal		= "Nat. Mater.",
  volume		= "23",
  pages			= "790--795",
  year			= "2024",
  doi			= "10.1038/s41563-024-01852-w"
}

@article{Grishunin2017,
  author		= "Kirill A. Grishunin and Nikita A. Ilyin and Natalia E. Sherstyuk and Elena D. Mishina and Alexey Kimel and Vladimir M. Mukhortov and Andrey V. Ovchinnikov and Oleg V. Chefonov and Mikhail B. Agranat",
  title ="{THz Electric Field-Induced Second Harmonic Generation in Inorganic Ferroelectric}",
  journal		= "Sci. Rep.",
  volume		= "7",
  pages			= "687",
  year			= "2017",
  doi			= "10.1038/s41598-017-00704-9"
}

@article{VonHoegen2018,
  author		= "A. von Hoegen and R. Mankowsky and M. Fechner and M. Först and A. Cavalleri",
  title ="Probing the interatomic potential of solids with strong-field nonlinear phononics",
  journal		= "Nature",
  volume		= "555",
  pages			= "79–-82",
  year			= "2018",
  doi			= "doi:10.1038/nature25484"
}

@article{b4,
  author		= "Ellen H. G. Backus and Nuria Garcia-Araez and Mischa Bonn and Huib J. Bakker",
  title ="On the Role of Fresnel Factors in Sum-Frequency Generation Spectroscopy of Metal–Water and Metal-Oxide–Water Interfaces",
  journal		= "J. Phys. Chem. C",
  volume		= "116",
  pages			= "23351-–23361",
  year			= "2012",
  doi			= "doi:10.1021/jp306273d"
}

@article{LiarXiv,
  author		= "Wei Li and Sixu Wang and Pai Peng and Haojie Han and Xinbo Wang and Jing Ma and Jianlin Luo and Jun-Ming Liu and Jing-Feng Li and Ce-Wen Nan and Qian Li",
  title ="Terahertz excitation of collective dynamics of polar skyrmions over a broad temperature range",
  journal		= "Nat. Phys.",
  year ="2025",
  doi ="10.1038/s41567-025-03056-8"
 }

@article{DanielBFO,
  author		= "Daniel A. Bustamante Lopez and Dominik M. Juraschek and Michael Fechner and Xianghan Xu and Sang-Wook Cheong and Wanzheng Hu",
  title ="Ultrafast simultaneous manipulation of multiple ferroic orders through nonlinear phonon excitation",
  journal		= "npj quantum materials",
  volume		= "10",
  pages			= "24",
  year			= "2025",
  doi			= "doi:10.1038/s41535-025-00738-7"
}

@article{Freire1988,
  author		= "J. D. Freire and R. S. Katiyar",
  title ="{Lattice dynamics of crystals with tetragonal BaTiO$_3$ structure}",
  journal		= "Phys. Rev. B ",
  volume		= "37",
  pages			= "2074--2085",
  year			= "1988",
  doi			= "doi:10.1103/PhysRevB.37.2074"
}

@article{HugoSubmitted,
  author		= "Huaiyu (Hugo) Wang and Vladimir Stoica and Cheng Dai and Marek Paściak and Sujit Das and Tiannan Yang and Mauro A. P. Gonçalves and Jiri Kulda and Margaret R. McCarter and Anudeep Mangu and Yue Cao and Hari Padma and Utkarsh Saha and Diling Zhu and Takahiro Sato and Sanghoon Song and Mathias Hoffmann and Patrick Kramer and Silke Nelson and Yanwen Sun and Quynh Nguyen and Zhan Zhang and Ramamoorthy Ramesh and Lane Martin and Aaron M. Lindenberg and Long-Qing Chen and John W. Freeland and Jirka Hlinka and Venkatraman Gopalan and Haidan Wen",
  title ="Ultrafast excitation of polar skyrons",
  journal		= "Nat. Commun.",
  volume		= "16",
  pages			= "8994",
  year ="2025",
  doi ="10.1038/s41467-025-64033-6"
 }

@article{Han2022,
  author		= "Lu Han and Christopher Addiego and Sergei Prokhorenko and Meiyu Wang and Hanyu Fu and Yousra Nahas and Xingxu Yan and Songhua Cai and Tianqi Wei and Yanhan Fang and Huazhan Liu and Dianxiang Ji and Wei Guo and Zhengbin Gu and Yurong Yang and Peng Wang and Laurent Bellaiche and Yanfeng Chen and Di Wu and Yuefeng Nie and Xiaoqing Pan",
  title ="High-density switchable skyrmion-like polar nanodomains integrated on silicon",
  journal		= "Nature",
  volume		= "603",
  pages			= "63–-67",
  year			= "2022",
  doi			= "10.1038/s41586-021-04338-w"
}

@article{Shkurinov1993,
  author		= "A. P. Shkurinov and A. V. Dubrovskii and N. I. Koroteev",
  title ="Second harmonic generation in an optically active liquid: Experimental observation of a fourth-order optical nonlinearity due to molecular chirality",
  journal		= "Phys. Rev. Lett.",
  volume		= "70",
  pages			= "1085--1088",
  year			= "1993",
  doi			= "10.1103/PhysRevLett.70.1085"
}

@article{Zhu2022,
  author		= "Ruixue Zhu and Zhexin Jiang and Xinxin Zhang and Xiangli Zhong and Congbing Tan and Mingwei Liu and Yuanwei Sun and Xiaomei Li and Ruishi Qi and Ke Qu and Zhetong Liu and Mei Wu and Mingqiang Li and Boyuan Huang and Zhi Xu and Jinbin Wang and Kaihui Liu and Peng Gao and Jie Wang and Jiangyu Liand Xuedong Bai",
  title ="Dynamics of Polar Skyrmion Bubbles under Electric Fields",
  journal		= "Phys. Rev. Lett.",
  volume		= "129",
  pages			= "107601 ",
  year			= "2022",
  doi			= "10.1103/PhysRevLett.129.107601"
}

@article{Liberal2017,
  author		= "Iñigo Liberal and Nader Engheta",
  title ="Near-zero refractive index photonics",
  journal		= "Nat. Photonics",
  volume		= "11",
  pages			= "149--158",
  year			= "2017",
  doi			= "10.1038/nphoton.2017.13"
}

@article{Reshef2019,
  author		= "Orad Reshef and Israel De Leon and M. Zahirul Alam and Robert W. Boyd",
  title ="Nonlinear optical effects in epsilon-near-zero media",
  journal		= "Nat. Rev. Mater.",
  volume		= "4",
  pages			= "535--551",
  year			= "2019",
  doi			= "10.1038/s41578-019-0120-5"
}

@article{Kinsey2019,
  author		= "Nathaniel Kinsey and Clayton DeVault and Alexandra Boltasseva and
Vladimir M. Shalaev",
  title = "Near-zero-index materials for photonics",
  journal		= "Nat. Rev. Mater.",
  volume		= "4",
  pages			= "742–-760",
  year			= "2019",
  doi			= "10.1038/s41578-023-00598-2"
}

@article{Hong2018,
  author		= "Zijian Hong and Long-Qing Chen",
  title = "Blowing polar skyrmion bubbles in oxide superlattices",
  journal		= "Acta Mater.",
  volume		= "152",
  pages			= "155--161",
  year			= "2018",
  doi			= "10.1016/j.actamat.2018.04.022"
}

@article{Behera2022,
  author		= "Piush Behera and Molly A. May and Fernando Gómez-Ortiz and Sandhya Susarla and Sujit Das and Christopher T. Nelson and Lucas Caretta and Shang-Lin Hsu and Margaret R. McCarter and Benjamin H. Savitzky and Edward S. Barnard and Archana Raja and Zijian Hong and Pablo García-Fernandez and Stephen W. Lovesey and Gerrit van der Laan and Peter Ercius and Colin Ophus and Lane W. Martin and Javier Junquera and Markus B. Raschke and Ramamoorthy Ramesh",
  title = "Electric field control of chirality",
  journal		= "Sci. Adv.",
  volume		= "8",
  pages			= "eabj8030",
  year			= "2022",
  doi			= "10.1126/sciadv.abj8030"
}

@article{Gao2024,
  author		= "Lingyuan Gao and Sergei Prokhorenko and Yousra Nahas and Laurent Bellaiche",
  title = "Dynamical Control of Topology in Polar Skyrmions via Twisted Light",
  journal		= "Phys. Rev. Lett.",
  volume		= "132",
  pages			= "026902",
  year			= "2024",
  doi			= "10.1103/PhysRevLett.132.026902"
}

@article{Nahas2020,
  author		= "Y. Nahas and S. Prokhorenko and Q. Zhang and V. Govinden and N. Valanoor and L. Bellaiche ",
  title = "Topology and control of self-assembled domain patterns in low-dimensional ferroelectrics",
  journal		= "Nature Commun.",
  volume		= "11",
  pages			= "5779",
  year			= "2020",
  doi			= "10.1038/s41467-020-19519-w"
}

@article{a1a2,
  author		= "Daniel A. Bustamante Lopez and Deepankar Sri Gyan and Sujit
Das and Michael Fechner and Xiaojiang Li and Lane W. Martin and Ramamoorthy Ramesh and Venkatraman Gopalan and Wanzheng Hu",
  journal		= "Submitted",
  title = "Phononic excitation of polar collective modes",
  year ="2025"
 }

@article{Wang2020,
  author		= "Y. J. Wang and Y. P. Feng and Y. L. Zhu and Y. L. Tang and L. X. Yang and M. J. Zou and W. R. Geng and M. J. Han and X. W. Guo and B. Wu and  X. L. Ma ",
  title = "Polar meron lattice in strained oxide ferroelectrics",
  journal		= "Nature Mater.",
  volume		= "19",
  pages			= "881–-886 ",
  year			= "2020",
  doi			= "10.1038/s41563-020-0694-8"
}

@article{Dutt2024,
  author		= "Avik Dutt and Aseema Mohanty and Alexander L. Gaeta and Michal Lipson",
  title = "Nonlinear and quantum photonics using integrated optical materials",
  journal		= "Nat. Rev. Mater.",
  volume		= "9",
  pages			= "321–-346",
  year			= "2024",
  doi			= "doi.org/10.1038/s41578-024-00668-z"
}

@article{Elshaari2020,
  author		= "Ali W. Elshaari and Wolfram Pernice and Kartik Srinivasan and Oliver Benson and Val Zwiller",
  title = "Hybrid integrated quantum photonic circuits",
  journal		= "Nat. Photonics",
  volume		= "14",
  pages			= "285–-298",
  year			= "2020",
  doi			= "doi.org/10.1038/s41566-020-0609-x"
}

@article{Chang2014,
  author		= "Darrick E. Chang and Vladan Vuletić and Mikhail D. Lukin",
  title = "Quantum nonlinear optics — photon by photon",
  journal		= "Nat. Photonics",
  volume		= "8",
  pages			= "685-–694",
  year			= "2014",
  doi			= "doi.org/10.1038/nphoton.2014.192"
}

@article{Dogadov2022,
  author		= "Oleg Dogadov and Chiara Trovatello and Baicheng Yao and Giancarlo Soavi and Giulio Cerullo",
  title = "Parametric Nonlinear Optics with Layered Materials and Related Heterostructures",
  journal		= "Laser Photonics Rev.",
  volume		= "16",
  pages			= "2100726",
  year			= "2022",
  doi			= " doi.org/10.1002/lpor.202100726"
}

@article{Akamatsu2018,
  author		= "H. Akamatsu and Y.K. Yuan and V.A. Stoica and G. Stone and T.N. Yang and Z.J. Hong and S.M. Lei and Y. Zhu and R.C. Haislmaier and J.W. Freeland and L.Q. Chen and H.D. Wen and V. Gopalan",
  title = "Light-Activated Gigahertz Ferroelectric Domain Dynamics",
  journal		= "Phys. Rev. Lett.",
  volume		= "120",
  pages			= "096101",
  year			= "2018",
  doi			= " doi.org/10.1103/PhysRevLett.120.096101"
}

@article{Yang2020,
  author		= "T.N. Yang and B. Wang and J.M. Hu and L.Q. Chen",
  title = "Domain Dynamics under Ultrafast Electric-Field Pulses",
  journal		= "Phys. Rev. Lett.",
  volume		= "124",
  pages			= "107601",
  year			= "2020",
  doi			= "doi.org/10.1103/PhysRevLett.124.107601"
}

@article{Chen1998,
  author		= "L.Q. Chen and J. Shen",
  title = "Applications of Semi-Implicit Fourier-Spectral Method to Phase Field Equations",
  journal		= "Computer Physics Communications",
  volume		= "108",
  pages			= "147--158",
  year			= "1998",
  doi			= "doi.org/10.1016/S0010-4655(97)00115-X"
}

@article{Tan2021,
  author		= "Congbing Tan and Yongqi Dong and Yuanwei Sun and Chang Liu and Pan Chen and Xiangli Zhong and Ruixue Zhu and Mingwei Liu and Jingmin Zhang and Jinbin Wang and Kaihui Liu and Xuedong Bai and Dapeng Yu and Xiaoping Ouyang and Jie Wang and Peng Gao and Zhenlin Luo and Jiangyu Li",
  title = "Engineering polar vortex from topologically trivial domain architecture",
  journal		= "Nat. Commun.",
  volume		= "12",
  pages			= "4620",
  year			= "2021",
  doi			= "doi.org/10.1038/s41467-021-24922-y"
}

@article{Haun1987,
  author		= "M. J. Haun and E. Furman and S. J. Jang and H. A. McKinstry and L. E. Cross",
  title = "Thermodynamic theory of {P}b{T}iO$_3$ ",
  journal		= "J. Appl. Phys.",
  volume		= "62",
  pages			= "3331–-3338",
  year			= "1987",
  doi			= "doi.org/10.1063/1.339293"
}

@article{Sheng2010,
  author		= "G. Sheng and Y. L. Li and J. X. Zhang and S. Choudhury and Q. X. Jia and V. Gopalan and D. G. Schlom and Z. K. Liu and L. Q. Chen",
  title = "A modified Landau–Devonshire thermodynamic potential for strontium titanate",
  journal		= "Appl. Phys. Lett.",
  volume		= "96",
  pages			= "232902",
  year			= "2010",
  doi			= "doi.org/10.1063/1.3442915"
}

@article{Luk2015,
  author		= "Ting S. Luk and Domenico de Ceglia and Sheng Liu and Gordon A. Keeler and Rohit P. Prasankumar and Maria A. Vincenti and Michael Scalora and Michael B. Sinclair and Salvatore Campione",
  title = "Enhanced third harmonic generation from the epsilon-near-zero modes of ultrathin films",
  journal		= "Appl. Phys. Lett.",
  volume		= "106",
  pages			= "151103",
  year			= "2015",
  doi			= "dx.doi.org/10.1063/1.4917457]"
}
